\documentclass[a4paper,11pt]{article}
\pdfoutput=1 

\usepackage{jcappub} 

\usepackage[T1]{fontenc} 
\usepackage{mathtools}

\def\lsim{\mathrel{\rlap{\lower4pt\hbox{\hskip1pt$\sim$}} \raise1pt\hbox{$<$}}}
\def\gsim{\mathrel{\rlap{\lower4pt\hbox{\hskip1pt$\sim$}} \raise1pt\hbox{$>$}}}
\usepackage{hyperref}
\usepackage{graphicx, color}
\usepackage[export]{adjustbox}

\newcommand{\newc}{\newcommand}
\newc{\comment}[1]{}
\newc{\dyc}{\textcolor{cyan}}

\title{The Inflaton that Could : Primordial Black Holes and Second Order Gravitational Waves from Tachyonic Instability induced in Higgs-$R^2$ Inflation}

\author[a]{Dhong Yeon Cheong,}

\affiliation[a]{Department of Physics \& IPAP \& Lab for Dark Universe, Yonsei University, Seoul 03722, Republic of Korea}

\emailAdd{dhongyeon@yonsei.ac.kr}

\author[b,c,d]{Kazunori Kohri,}

\affiliation[b]{Theory Center, IPNS, KEK, 1-1 Oho, Tsukuba, Ibaraki 305-0801, Japan}
\affiliation[c]{The Graduate University for Advanced Studies (SOKENDAI), 1-1 Oho, Tsukuba, Ibaraki 305-0801, Japan}
\affiliation[d]{Kavli IPMU (WPI), UTIAS, The University of Tokyo, Kashiwa, Chiba 277-8583, Japan}

\emailAdd{kohri@post.kek.jp}

\author[a,e]{Seong Chan Park}

\affiliation[e]{Korea Institute for Advanced Study, Seoul 02455, Republic of Korea}

\emailAdd{sc.park@yonsei.ac.kr}

\abstract{The running of the Higgs self coupling may lead to numerous phenomena in early universe cosmology. In this paper we introduce a scenario where the Higgs running induces turns in the trajectory passing a region with tachyonic mass, leading to a temporal tachyonic growth in the curvature power spectrum.  This effect induced by the Higgs leaves phenomena in the form of primordial black holes and stochastic gravitational waves, where proposed GW observatories will be able to probe in the near future. }

\begin{document}
\maketitle
\flushbottom

\section{Introduction}

We are now officially entering the era of gravitational wave (GW) observatories. After the discovery of GWs from LIGO/VIRGO, the database including gravitational wave signals of binary systems increased significantly~\cite{LIGOScientific:2016aoc, LIGOScientific:2021djp}. Recently, NANOGrav and several pulsar timing arrays reported a background which may represent a stochastic gravitational wave background~\cite{NANOGrav:2020bcs, Goncharov:2021oub,Wu:2021kmd,  NANOGrav:2021ini, Xue:2021gyq, Antoniadis:2022pcn}, where scenarios incorporate solar-mass primordial black holes \cite{Vaskonen:2020lbd, DeLuca:2020agl, Kohri:2020qqd,Domenech:2020ers}. Many proposed GW observatories (e.g. LISA~\cite{Baker:2019nia, LISACosmologyWorkingGroup:2022jok}, DECIGO~\cite{Sato:2017dkf, Kawamura:2020pcg}, Einstein Telescope~\cite{Maggiore:2019uih}, SKA~\cite{Weltman:2018zrl}, etc.) expect to cover a wide range of frequencies, further unraveling physics occurring in the early universe~{\cite{Sedda:2019uro, Caldwell:2022qsj}}.

Numerous scenarios in our early universe may produce stochastic GW backgrounds (SGWB), 
which include cases that induce stochastic GWs at the second order (a comprehensive review regarding this topic can be found in \cite{Domenech:2021ztg}). Processes induced by inflation gained much interest with a localized enhancement in the curvature power spectrum \cite{Ivanov:1994pa,Green:1997sz,Drees:2011hb,Kohri:2012yw,Clesse:2015wea, Nakama:2016gzw, Inomata:2017okj, Garcia-Bellido:2017mdw, Domcke:2017fix, Kannike:2017bxn, Carr:2017edp, Germani:2017bcs,Motohashi:2017kbs, Pattison:2017mbe, Di:2017ndc, Inomata:2017vxo,Ando:2017veq,Hertzberg:2017dkh,Franciolini:2018vbk,Biagetti:2018pjj,Cai:2018tuh,Germani:2018jgr,Dalianis:2018frf,Byrnes:2018txb,Passaglia:2018ixg,Dimopoulos:2019wew, Bhaumik:2019tvl,Carrilho:2019oqg,Fu:2019ttf,Mishra:2019pzq,Cai:2019bmk,Cheong:2019vzl, Ashoorioon:2019xqc, Lin:2020goi,  Ballesteros:2020qam, Palma:2020ejf,Fumagalli:2020adf, Braglia:2020eai, Aldabergenov:2020bpt, Aldabergenov:2020yok, Gundhi:2020kzm, Gundhi:2020zvb, DeLuca:2020agl,Kohri:2020qqd, Zheng:2021vda,  Chen:2021nio, Kawai:2021edk, Rezazadeh:2021clf,  Iacconi:2021ltm,  Pi:2021dft, Papanikolaou:2021uhe,Ashoorioon:2022raz, Kallosh:2022vha,Geller:2022nkr,Karam:2022nym}, in correlation with copious primordial black hole (PBH) production~\cite{Cai:2018dig, Bartolo:2018evs, Braglia:2020taf, Braglia:2021wwa}.\footnote{  For a recent review on PBHs see \cite{Carr:2020gox} and references within. }

Among many inflationary models, Higgs inflation~\cite{Bezrukov:2007ep} and its extensions~\cite{Giudice:2010ka,Bezrukov:2010jz,Lerner:2011it,Calmet:2013hia,Hamada:2014iga,Bezrukov:2014bra,Herranen:2014cua,Hamada:2014wna,Bezrukov:2014ipa,Barbon:2015fla,Salvio:2015kka,Calmet:2016fsr,Hamada:2016onh,Ema:2017rqn,Ezquiaga:2017fvi,Bezrukov:2017dyv,Hamada:2017sga,Lee:2018esk,He:2018gyf,Masina:2018ejw,Gorbunov:2018llf,Ghilencea:2018rqg,Gundhi:2018wyz,Rasanen:2018fom,He:2018mgb,Bezrukov:2019ylq,Drees:2019xpp,Ema:2019fdd,Cheong:2019vzl,Hamada:2020kuy,Ema:2020evi,Gundhi:2020zvb, Cheong:2021vdb, Lee:2021dgi, Aoki:2021aph,Lee:2021rzy,  Cheong:2021kyc, Aoki:2022dzd} gain immense interest as it incorporates the Standard Model scalar with a nonminimal coupling to gravity, and it provides the best fit to current cosmic microwave background (CMB) observations. The running behavior of the Standard Model Higgs is also incorporated in the potential, which prospects numerous phenomena in our cosmology~\cite{Hamada:2014iga, Hamada:2014wna, Hamada:2016onh, Hamada:2017sga}. A general setup incorporating this is the Higgs-$R^2$ inflation, where two scalars, namely the scalaron and the SM Higgs generate a two field potential~\cite{Salvio:2015kka,Calmet:2016fsr,Wang:2017fuy,Ema:2017rqn,Pi:2017gih,Ghilencea:2018rqg, Gorbunov:2018llf,He:2018gyf, He:2018mgb, Gundhi:2018wyz, Ema:2019fdd, Bezrukov:2019ylq, Cheong:2019vzl, Ema:2020evi, He:2020ivk, Bezrukov:2020txg, He:2020qcb, Gundhi:2020zvb,  Lee:2021dgi, Aoki:2021aph, Aoki:2022dzd}.  

Intriguingly, the running of the Higgs self coupling running can induce much richer phenonema. We discussed the parameters that induce an inflection point in the model describable in the framework of an effective single field case in our previous work \cite{Cheong:2019vzl}. There, we concluded that an enhanced curvature perturbation can be produced by a near-inflection point induced by the Higgs running, resulting in a tight correlation with the PBH mass and the CMB spectral index. In this paper, we revisit the Higgs running and show cases where the inflaton possesses turns in its trajectory and approaches the hill in the potential at $h=0$, which exhibits a tachyonic mass.\footnote{Previous studies on the tachyonic instability in Higgs-$R^2$ focused on the preheating era~\cite{ Bezrukov:2019ylq,  He:2020ivk, Bezrukov:2020txg}. } Isocurvature perturbations grow exponentially, which induce a rapid growth of the curvature perturbations (this mechanism, mainly incorporating a rapid turn in the non-geodesic field space has taken interest in the past several years~\cite{Palma:2020ejf, Fumagalli:2020adf, Braglia:2020eai, Gundhi:2020kzm, Gundhi:2020zvb, Fumagalli:2020nvq, Kallosh:2022vha, Boutivas:2022qtl}). This in turn displays a sharp bump in the curvature power spectrum, where this local feature is probe-able in the form of PBHs and stochastic GWs in a wide range of masses and frequencies. The mass and abundance of the PBHs, and correspondingly the energy density and the frequency of the stochastic GWs depend on the parameter choices $(\xi, \lambda)$, which allows one to connect and probe low energy Standard Model measurements with proposed gravitational wave measurements. 

This paper is organized as follows, we introduce the Higgs-$R^2$ setup including the running behavior of the Higgs. We analyze the background dynamics of the inflaton and classify the trajectory in steps. We then compute the curvature and isocurvature perturbations of the model and its corresponding PBH abundance and GW spectrum. We conclude with the implications of the results.

\section{Inflation action}
\label{sec:action}

The action for the Higgs-$R^2$ inflation in the Jordan frame is given as 
\begin{equation}
S_J =  \int d^4 x \sqrt{-g_{J}} \left[ \frac{M_P^2}{2} \left(R_{J} + \frac{\xi(\mu)\, h^2}{M_P^2}R_J+ \frac{R_J^2}{6M^2} \right) - \frac{1}{2} g^{\mu\nu} \nabla_\mu h \nabla_\nu h - \frac{\lambda(\mu)}{4} h^4 \right], 
\label{eq:jordanaction}
\end{equation}
with the Higgs, $h$, in the unitary gauge, the reduced Planck mass being $M_P=1/\sqrt{8\pi G}\simeq 2.44\times 10^{18}~{\rm GeV}$, and the scalaron mass $M\lsim M_P/\xi$ introduced to match the dimensions. 
We take the Higgs self coupling running $\lambda\left(\mu\right)$ at a scale $\mu$. 
The scalaron, $s$,  is defined via
\begin{equation}
\sqrt{\frac{2}{3}} \frac{s}{M_{P}} = \ln \left(1 + \frac{\xi h^2}{M_P^2} + \frac{R_J}{3 M^2}\right) \equiv \Omega(s).
\label{eq:scalarondef}
\end{equation}
Weyl transformation yields the action in Einstein frame where  $g_{\mu\nu} =e^{\Omega(s)} g^J_{\mu\nu}$ with two scalar fields, $(\phi^a)=(s,h)$ appearing in the scalar potential $U(\phi^a)$. As a consequence, the kinetic terms involve a nontrivial field space metric $G_{ab}$:
\begin{align}
S= \int d^4 x \sqrt{-g}\left[ \frac{M_P^2}{2}R  - \frac{1}{2}G_{ab} g^{\mu \nu} \nabla_\mu \phi^a \nabla_\nu \phi^b - U(\phi^a) \right], \label{eq:einsteinaction} \\
U(\phi^a) \equiv e^{-2\Omega(s)} \left\{ \frac{3}{4}M_P^2 M^2  \left(e^{\Omega(s)} - 1 - \frac{\xi(\mu) h^2}{M_P^2}\right)^2  +\frac{\lambda\left(\mu\right)}{4}h^4 \right\}.
\label{eq:einsteinpotential}
\end{align}
Explicitly, the field space metric is given for $(s,h)$ as 
\begin{equation}
G_{ab}=\begin{pmatrix}
1&&0\\
0&&e^{-\Omega(s)}
\end{pmatrix}.
\label{eq:fieldspacemetric}
\end{equation}

The parameters $\left.(M, \xi, \lambda)\right|_\mu$ running in scale $\mu$ by the Standard Model and scalaron interactions follow 1-loop beta functions in the form \cite{Muta:1991mw, Elizalde:1993ee, Elizalde:1993ew, Codello:2015mba, Markkanen:2018bfx, Gorbunov:2018llf, Ema:2019fdd,Ema:2020evi}
{\begin{align}
\beta_\alpha &= -\frac{1}{16\pi^2}\frac{\left(1+6\xi\right)^2}{18} ,\\
\beta_\xi &= + \frac{1}{16\pi^2 }\left(\xi + \frac{1}{6}\right)\left(12 \lambda + 6 y_t^2 - \frac{3}{2}g'^2 - \frac{9}{2}g^2\right) ,\\
\beta_\lambda &= \beta_\text{SM} + \frac{1}{16\pi^2} \frac{2\xi^2 \left(1+6\xi\right)^2 M^4}{M_P^4},
\end{align}
}
with $\alpha = {M_P^2}/{12 M^2}$ and $\beta_{\rm SM}$ being the Standard Model contribution~\cite{DeSimone:2008ei}. 
{ 
Choosing the renormalization prescription to be $\mu \simeq \sqrt{h^2}$, we perform a standard parameterization of the parameters $\lambda (\mu)\, ,\,\, \xi (\mu) $~\footnote{The running of $\alpha$ with the parameters considered in this paper are characterized as $|\beta_\alpha | \sim \mathcal{O}(1)$. Note that this is infinitesimal to typical $\alpha \simeq \mathcal{O} (10^{10}) $ values needed for successful inflation. Therefore we safely neglect its running effects and take $\alpha$ as a constant.} around the $\lambda$ minimum field value $h_m$ 
\begin{align}
\left. \lambda\left(\mu \right) \right|_{\mu = h} & = \lambda_m + \frac{\beta_2^\text{SM}}{\left(16 \pi^2\right)^2} \ln^2\left(\sqrt{\frac{h^2}{h_m^2}}\right) = \lambda_m + b \ln^2 \left(\sqrt{\frac{h^2}{h_m^2}}\right)
\label{eq:lambdarunning} \\ 
\left. \xi \left(\mu \right) \right|_{\mu = h} &= \xi_0 + 2{\beta_\xi^0} \ln \left(\sqrt{\frac{h^2}{h_m^2}}\right) = \xi_0 + b_\xi \ln  \left(\sqrt{\frac{h^2}{h_m^2}}\right)
\label{eq:xirunning} 
\end{align}
with $\lambda_m \equiv \lambda (h_m)\sim \mathcal{O}(10^{-6})  $, $\xi_0 \equiv \xi ( h_m) \sim \mathcal{O}(1) $ ,  $\beta_2^{\text{SM}} \sim 0.5$, $ \beta_\xi^0 \equiv \beta_\xi(h_m) \sim - 0.01$, $\mu_{m} = h_{m} \sim 10^{17} - 10^{18}\,\, \text{GeV}$ as denoted in  \cite{Degrassi:2012ry, Buttazzo:2013uya}. \footnote{We take $ \lambda_{m} > 0 $ to guarantee the stability of the Higgs potential during inflation. 
} 
The 1-loop $\beta$-functions indicate that the running effects are most significant in $\lambda$, with many orders changing while running from EW scales to Planck scales, whereas the $\xi$ parameter running is insignificant over this running range maintaining the same order.  
Throughout the paper we take $M_P=1$  and focus on the $h>0$ region unless specified.} 

\section{Background evolution}
The action eq.~(\ref{eq:einsteinaction}) yields the equations of motion for the homogeneous background fields and the Friedmann equation with the metric~\cite{He:2018gyf}
\begin{align}
ds^2 = -dt^2 +a(t)^2 \delta_{ij} dx^i dx^j
\end{align}
gives,  expressed incorporating the `curved field space metric' effects,
\begin{align}
D_t \dot{\phi}^a + 3 H \dot{\phi}^a + G^{ab}D_b U & =0, \label{eq:eom}\\
3H^2 &= \frac{1}{2}\dot{\phi}_0^2 + U 
\end{align}
with the covariant derivatives  $D_a \phi^b  = \partial_a \phi^b  + \Gamma^b_{c a} \phi^c$, $\Gamma^b_{ca} = \frac{1}{2} G^{be}\left(\partial_c G_{ae } + \partial_a G_{ec} - \partial_{e} G_{ca }\right)$, $D_t X^a  = \dot{X}^a + \Gamma_{bc}^a \dot{\phi}^{b}X^c$, and $\dot{\phi}_0^2 = G_{ab}\dot{\phi}^a \dot{\phi}^b$.  

\begin{figure}[t!]
   \centering
   \includegraphics[width=.7\textwidth]{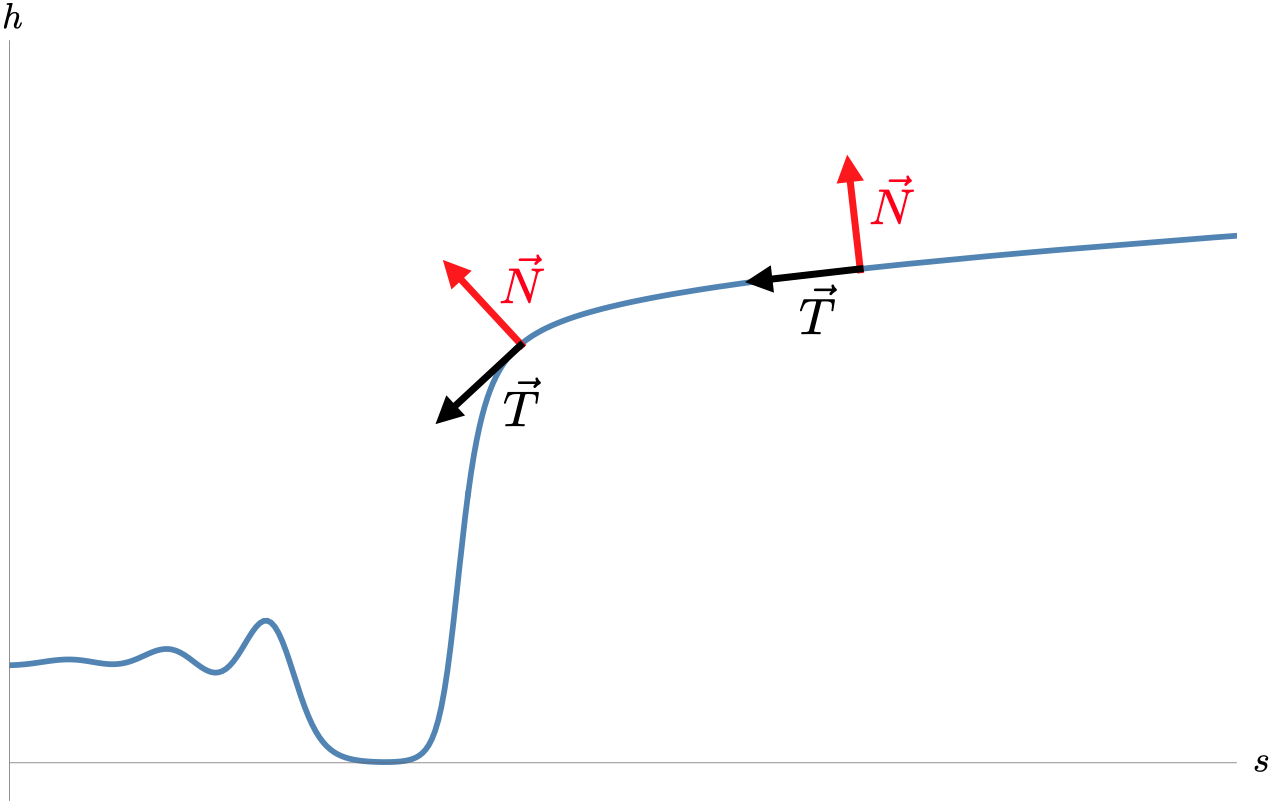} 
   \caption{Tangent and normal vectors $\vec{T}$, $\vec{N}$ in field space $(s, h)$. }
   \label{fig:fieldvectors}
\end{figure}

The trajectory then takes a unique path in the field space $(s, h)$. The parameterization of this curve can be described by constructing a set of orthogonal unit vectors $T^a(t)$ and $N^a(t)$ where the former is tangent to the path and the latter is normal to it, as depicted in figure~\ref{fig:fieldvectors}. Explicitly,
\begin{equation}
T^a = \frac{\dot{\phi}^a}{\dot{\phi}_0 } \,\, , \,\,\,\, N_a = \sqrt{\text{det}{\,G}} \,\epsilon_{ab}T^b
\end{equation}
with $\epsilon_{ab}$ being the 2 dimensional Levi-Civita symbol.
Projection of the equations of motion to the tangent vector 
\begin{equation}
\ddot{\phi}_ 0 + 3H \dot{\phi}_0 + U_T = 0
\end{equation}
with $U_T \equiv T^a U_a$. Projections to the normal vector $N^a$ gives
\begin{equation}
\frac{DT^a}{dt} = - \frac{U_N}{\dot{\phi}_0} N^a .
\label{eq:normalprojection}
\end{equation}
We now define the slow-roll parameters by generalizing the setup to a multifield scenario. They take the form
\begin{equation}
\epsilon \equiv -\frac{\dot{H}}{H^2} = \frac{\dot{\phi}_0^2}{2 H^2} \,\,,\,\,\,\, 
\eta^a \equiv - \frac{1}{H \dot{\phi}_0}D_t \dot{\phi}^a .
\end{equation}
Note that, the $\eta^a $ parameter is now a vector in the sense that there are two degrees in the field space. The same decomposition to $T^a$ and $N^a$ can be performed to this parameter as well, leading to 
\begin{equation}
\eta^a = \eta_{\parallel} T^a + \eta_{\perp} N^a 
\end{equation}
with 
\begin{align}
\eta_\parallel \equiv - \frac{\ddot{\phi}_0}{H \dot{\phi}_0}
 \,\,,\,\,\,\, 
\eta_\perp \equiv  \frac{U_N}{\dot{\phi}_0 H }. \end{align}
Given that $\eta_\parallel$ is along the tangential direction of the trajectory, it can be regarded as the extension to the normal $\eta$ slow roll parameter in single field inflation. $\eta_\perp$ on the other hand, can also be inserted into eq.~(\ref{eq:normalprojection}) in the sense that
\begin{equation}
\frac{D T^a}{dt} = - H \eta_\perp N^a \equiv - \dot{\theta} N^a
\end{equation}
hence, the $\eta_\perp$ parameter precisely shows how quickly the tangential direction $T^a$ is varying in time. The two parameters are related as $\dot{\theta} \equiv H \eta_\perp$.

\section{Cosmological perturbations}
Having the background evolutions, we now perturb the action and describe the scalar perturbations of the model. The notations used are based on \cite{Cespedes:2012hu, Achucarro:2012yr} (see also \cite{GrootNibbelink:2001qt, DiMarco:2002eb, Peterson:2011yt, Greenwood:2012aj, Kaiser:2013sna}). 

The fields $\phi^a$ and the metric can be perturbed as 
\begin{align}
\phi^a(t, \vec{x}) &= \phi_0^a (t) + \delta \phi^a (t,\vec{x}), \\
ds^2 &= -(1+2 \psi) dt^2 + a(t)^2(1-2\psi)\delta_{ij} dx^i dx^j . 
\end{align}
Implementing the basis $T^a$ and $N^a$ to the perturbations allows the following gauge invariant fields
\begin{align}
v_T &= a T_a \delta \phi^a + a \frac{\dot{\phi}_0}{H} \psi  \equiv a T_a Q^a\\
v_N &= a N_a \delta \phi^a \equiv a N_a Q^a
\end{align}
with $Q^a \equiv \delta \phi^a + \frac{\dot{\phi}^a}{H}\psi$ being the Mukhanov-Sasaki variable. In terms of these variables we also define the comoving curvature/isocurvature perturbation 
\begin{align}
\mathcal{R} = \frac{H}{a \dot{\phi}_0} v_T \equiv \frac{H}{\dot{\phi}_0} Q_T \\
\mathcal{S} = \frac{H}{a \dot{\phi}_0} v_N \equiv \frac{H}{\dot{\phi}_0} Q_N .
\end{align}
The perturbed action up to second order is then
\begin{equation}
S^{(2)} = \frac{1}{2} \int d^4 x a^3 \left[ \frac{\dot{\phi}_0^2}{H^2} \dot{\mathcal{R}}^2  - \frac{\dot{\phi}_0^2}{H^2} \frac{(\nabla \mathcal{R})^2}{a^2} + {\dot{Q}_N}^2 -  \frac{(\nabla Q_N)^2}{a^2} + 4 \dot{\phi}_0 \eta_\perp \dot{\mathcal{R}} Q_N - M_\text{eff}^2 Q_N^2 \right]
\end{equation}
where $M_\text{eff}^2$ is
\begin{equation}
M_\text{eff}^2  = U_{NN} + H^2 \epsilon \mathbb{R} - \dot{\theta}^2 .
\label{eq:isocurvaturemass}
\end{equation}
The equations of motion are
\begin{align}
&\ddot{\mathcal{R}} + \left(3 + 2\epsilon - 2 \eta_{\parallel} \right)H \dot{\mathcal{R}} + \frac{k^2}{a^2} \mathcal{R} = -2 \frac{H^2}{\dot{\phi}_0} \eta_\perp \left[\dot{Q}_N + \left(3 - \eta_\parallel + \frac{\dot{\eta}_\perp}{H \eta_\perp}\right) H Q_N  \right] 
\label{eq:adiabaticpert}\\
&\ddot{Q}_N + 3 H \dot{Q}_N + \left(\frac{k^2}{a^2}+ M_\text{eff}^2 \right)Q_N = 2 \dot{\phi}_0 \eta_\perp \dot{\mathcal{R}} . 
\label{eq:isocurvaturepert}
\end{align}
Both eq.~(\ref{eq:adiabaticpert}) and eq.~(\ref{eq:isocurvaturepert}) incorporate mixing between $\mathcal{R}$ and $Q_N$, with the mixing proportional to $\eta_\perp$. A naive estimation yields when $\eta_{\perp} \gsim 1 \rightarrow \dot{\theta} \gsim H$, mixing between the two perturbations become significant and $Q_N$ can source $\mathcal{R}$.

In addition to the mixing, eq.~(\ref{eq:isocurvaturemass}) incorporates the essentials that determine the dynamics of $Q_N$. $M_\text{eff}^2$ can take a negative value either through 1) $U_{NN} <0 $, 2) $\mathbb{R} <0 $, corresponding to a hyperbolic geometry in the field space, 3) $\dot{\theta}^2 > U_{NN}$. In any case, a tachyonic isocurvature mass then modifies the equations of motion for $k^2/a^2 < |M_\text{eff}^2|$ to be
\begin{equation}
\ddot{Q}_N + 3 H \dot{Q}_N  -\left( \left|M_\text{eff}^2\right| \right)Q_N \simeq 0 . 
\end{equation}
Hence $Q_N$ can exhibit an \textit{exponential} growth due to the tachyonic mass. This growth can be more rapid than cases implementing a USR phase.

\section{Inflaton and perturbation evolution}
\begin{figure}[t]
   \centering
   \includegraphics[width=.9\textwidth]{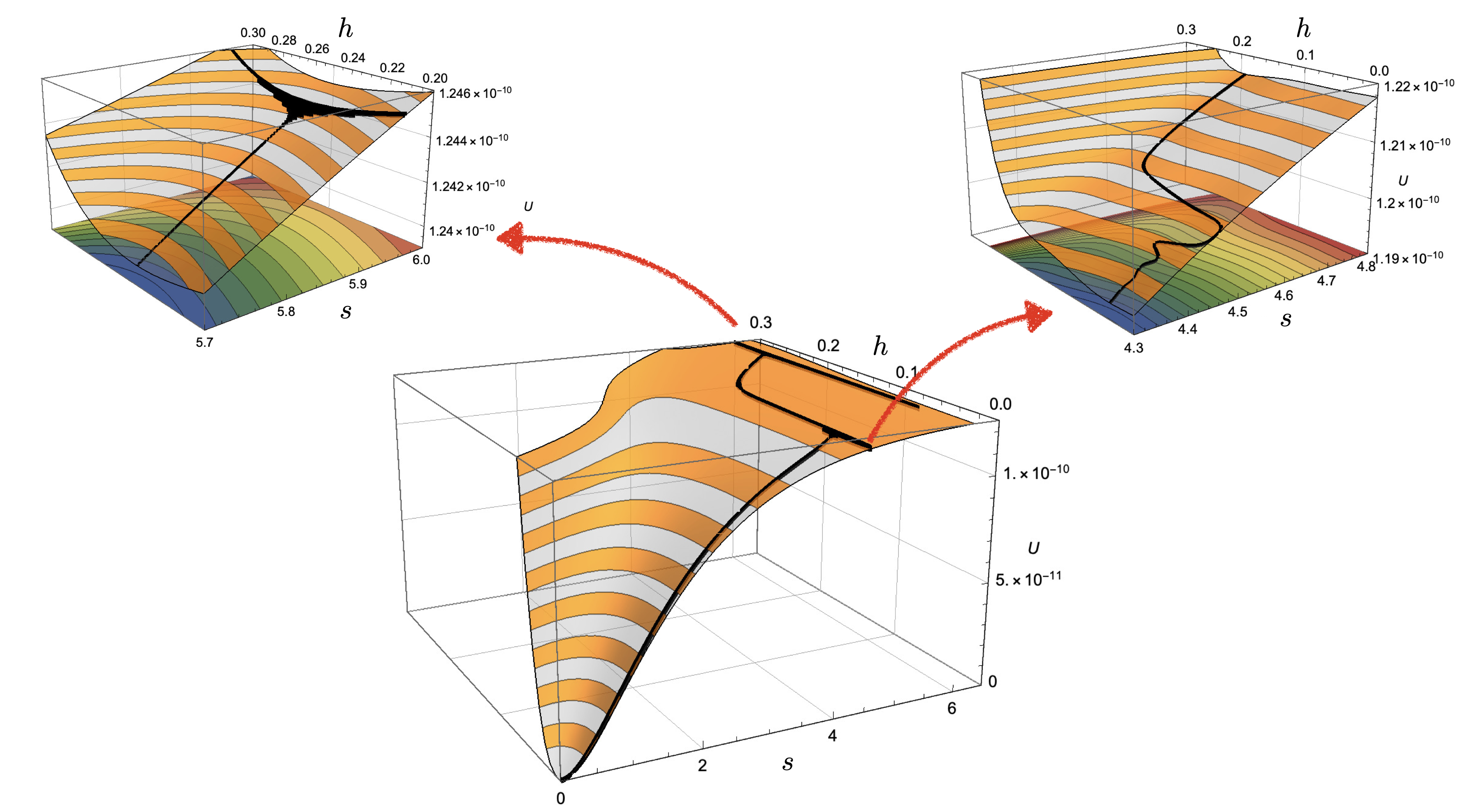} 
   \caption{Potential and field trajectory (black) of the setup. Note that the inflaton follows a well defined valley with a positive $M_\text{eff}^2$, allowing the trajectory to be independent to the initial conditions (top-left). The potential then exhibits a region where the valley disappears, inducing a turn in the trajectory and approaches the hill (top-right). After evolving along this tachyonic hill, the inflaton falls back down into a valley.  }
   \label{fig:potentialandtrajectory}
\end{figure}

We now turn our interest to the dynamics in the critical Higgs-$R^2$ setup, starting with the trajectory, which is schematically depicted in figure~\ref{fig:potentialandtrajectory}. 

\begin{itemize}
	\item {\bf Stage 1:}  Initially the inflaton starts rolling down a well defined valley, satisfying the slow-roll conditions. The large and positive $M_\text{eff}^2$ ensures the isocurvature perturbation $Q_N$ be suppressed. \textit{This initial valley allows the large scale predictions (e.g. CMB) to be insensitive to the initial values of the inflaton, simply speaking it exhibits an attractor prediction.}
	
	\item {\bf Stage 2:} Then the inflaton rolls down in the Higgs direction, approaching the hill at $h = 0$.
	
	\item {\bf Stage 3:} Once the inflaton climbs up the hill where $h \approx 0$, the inflaton exhibits a turn, and now evolves in the $s$ direction. The isocurvature mass $M_\text{eff}^2$ becomes large and negative, with its value mainly determined by $\xi$. Its precise value takes the form
{
\begin{align}
M_\text{eff}^2 \simeq \frac{1}{\dot{s}^2 +e^{-\sqrt{\frac{2}{3}} {s}} \dot{h}^2 } \left(e^{\sqrt{\frac{2}{3}} {s}} \dot{s}^2 \frac{\partial^2 U}{\partial h^2}  \right) 
 \simeq -3 M^2  \left[ \xi_0 + b_\xi \ln \left(\frac{h}{h_m} \right) + \frac{9}{2} b_\xi  \right] \left(1-e^{-\sqrt{\frac{2}{3}}{s}}\right) . 
\label{eq:meffapprox}
\end{align}
}
The inflaton then exhibits another turn back into the valley. 

\item {\bf Stage 4:} The inflaton once again rolls into the well defined valley, with a large and positive $M_\text{eff}^2$. 
\end{itemize}
Let's look into Stage 1, 2, 3 in more detail.

\subsection{Stage 1. }
In this region, where $M_\text{eff}^2 > H^2 > 0$ and the turn rate $\dot{\theta} \ll H$, the equations of motion simply resemble standard \textit{effective single field, slow-roll} results. The equations take the form, with the transformation of the time variable to efolds using $N_e = \int^{t} dt^\prime H(t^\prime)$ 
\begin{align}
\frac{d^2 \mathcal{R}_k}{dN_e^2}  + \left(3 + \epsilon - 2 \eta_{\parallel} \right) \frac{d {\mathcal{R}_k}}{dN_e} + \frac{k^2}{a^2 H^2}  \mathcal{R} &= 0 \label{eq:adiabaticpertstage1}\\
\frac{d^2 {Q}_{N,k}}{dN_e^2} + 3\frac{d {Q}_{N,k}}{dN_e} + \left(\frac{k^2}{a^2 H^2}+ \frac{M_\text{eff}^2}{H^2} \right)Q_{N,k} &= 0 
\label{eq:isocurvaturepertstage1}
\end{align}
which, neglecting the slow-roll parameters as they are suppressed, give the generally known solutions with ${\epsilon_k^2 \equiv k^2 / a^2 H^2}$  \cite{Fumagalli:2020nvq, Boutivas:2022qtl}
\begin{align}
\mathcal{R}_k (N_e) &= e^{-\frac{3}{2} N_e}\left[c_1\,e^{-\frac{N_e}{2}  \sqrt{9 - 4 \epsilon_k^2 } } + c_2 \, e^{\frac{N_e}{2}  \sqrt{9 - 4 \epsilon_k^2 } }  \right]\label{eq:curvatureperturbationefold} \\ 
{Q}_{N, k} (N_e) & = e^{-\frac{3}{2} N_e}\left[c_3\,e^{-\frac{N_e}{2}  \sqrt{9  -4 \frac{M_\text{eff}^2}{H^2} -4 \epsilon_k^2 } } + c_4 \, e^{\frac{N_e}{2}  \sqrt{9  -4 \frac{M_\text{eff}^2}{H^2} -4 \epsilon_k^2  } }  \right]\label{eq:isocurvatureperturbationefold}
\end{align}
where  $c_1, \, c_2,\,  c_3, \,  c_4$ are determined by the conditions at the in-horizon state. We can see that for the case when $\epsilon_k^2 \ll 1$, i.e. out of the horizon, $\mathcal{R}_k (N_e) \propto \mathcal{R}_0 + \mathcal{R}_1 e^{-3N_e}$ and has a mode \textit{freezing out}, being constant deep outside the horizon. The isocurvature mode $Q_{N,k}$, in contrast, always is suppressed as $Q_{N,k}(N_e) \propto e^{-\frac{3}{2} N_e } e^{\pm i \frac{M_\text{eff}}{H}  N_e } $ due to $M_\text{eff} \gg H$ in this regime, and therefore is exponentially suppressed in the deep out of horizon region, giving negligible effects on cosmological observables. 

We now obtain standard effective single field slow-roll observables. {As the inflaton evolution in this era remains in the same $h$ order, we approximate $\xi(h) \simeq \xi_0$.  In this period, the fields resemble the approximate relation in the large-scale observable region}
\begin{equation}
s_v \approx \sqrt{\frac{3}{2}} \ln \left[1+ \frac{4(\lambda_m + 3 M^2 \xi_0^2 ) h^2 +(7h -5 h_m)(h - h_m) b  }{12 M^2 \xi_0} \right]
\end{equation}
allowing us to combine the fields to be parameterized with the scalaron only, leading to the slow roll parameters
\begin{equation}
\epsilon_H \simeq \epsilon_V \equiv \frac{1}{2} \left. \left[\frac{U_{s}(s, h(s))}{U(s, h(s))} \right]^2\right|_{s = s_*} \, \, , \, \, \,\,\,  \eta_{\parallel} \simeq \eta_V \equiv \left. \frac{U_{ss}(s, h(s))}{U(s, h(s))} \right|_{s = s_*}
\end{equation}
with $s_{*}$ being the scalaron field value at the CMB pivot scale and
\begin{equation}
n_s \equiv 1+ \frac{d \ln{\mathcal{P}_\mathcal{R}(k)}}{d \ln k} \simeq 1 - 6\epsilon_V + 2 \eta_V \, \, , \, \,\, \, \, r \simeq 16 \epsilon_V.
\end{equation}
 The inflationary epoch exhibits slow-roll with $\eta_H \ll 1$, and the $ \frac{\lambda(h)}{4} h^4  $ term in the potential eq.~\ref{eq:einsteinpotential} is orders smaller than other terms. This lets us reasonally take the inflationary efolds $N_\text{inf} = N_\text{end} - N_\text{pivot} \approx \frac{3}{4} e^{\sqrt{\frac{2} {3}} s_{*} }$, which resembles an $R^2$ inflation-like form. Therefore, the above expressions can be approximately expressed to
\begin{align}
n_s \approx &\,\,  1 - \frac{2}{N_\text{inf}} - \frac{9}{2 N_\text{inf}^2 }  \\ 
& +\frac{M^2 \xi_0^2\, b}{\lambda_m \left(\lambda_m + 3 M^2 \xi_0^2 \right)}\left[ 2 + \frac{3\left(2 + 3 \ln \left(\frac{(\lambda_{m} + 3 M^2 \xi_0^2 ) h_m^2 }{4 M^2 \xi_0 N_\text{inf}}\right) \right)}{2 N_\text{inf}} + \frac{27 \left(1 + \ln \left( \frac{(\lambda_m + 3 M^2 \xi_0^2 ) h_m^2 }{4 M^2 \xi_0 N_\text{inf}} \right)\right)}{4N_\text{inf}^2}\right] \nonumber
\end{align}  
and
\begin{equation}
\resizebox{\columnwidth}{!}{$r \approx \frac{12}{N_\text{inf}^2} + \frac{2 M^2 \xi_0^2 \, b}{\lambda_m (\lambda_m + 3M^2 \xi_0^2 ) N_\text{inf} } \left[12 \ln \left(  \frac{4M^2 \xi_0 N_\text{inf}  }{\left(\lambda_m + 3 M^2 \xi_0^2 \right) h_m^2 }\right)  + \frac{9 \left( 2 \ln\left( \frac{4 M^2 \xi_0  N_\text{inf}  }{\left(\lambda_m + 3 M^2 \xi_0^2  \right) h_m^2 } \right)- 1\right)}{ N_\text{inf}}\right] $ }
\end{equation}
where $N_\text{end}$ describes the efolds at the end of inflation, and $N_{\text{pivot}}$ represents the efolds at the CMB pivot scale. Therefore the additional logarithmic running of the Higgs self coupling \textit{shifts the spectral index  of the curvature power spectrum to larger values} compared to the pure Higgs-$R^2$ case with a constant Higgs self coupling. 

\begin{figure}[t!]
   \centering
   \includegraphics[width=.48\textwidth]{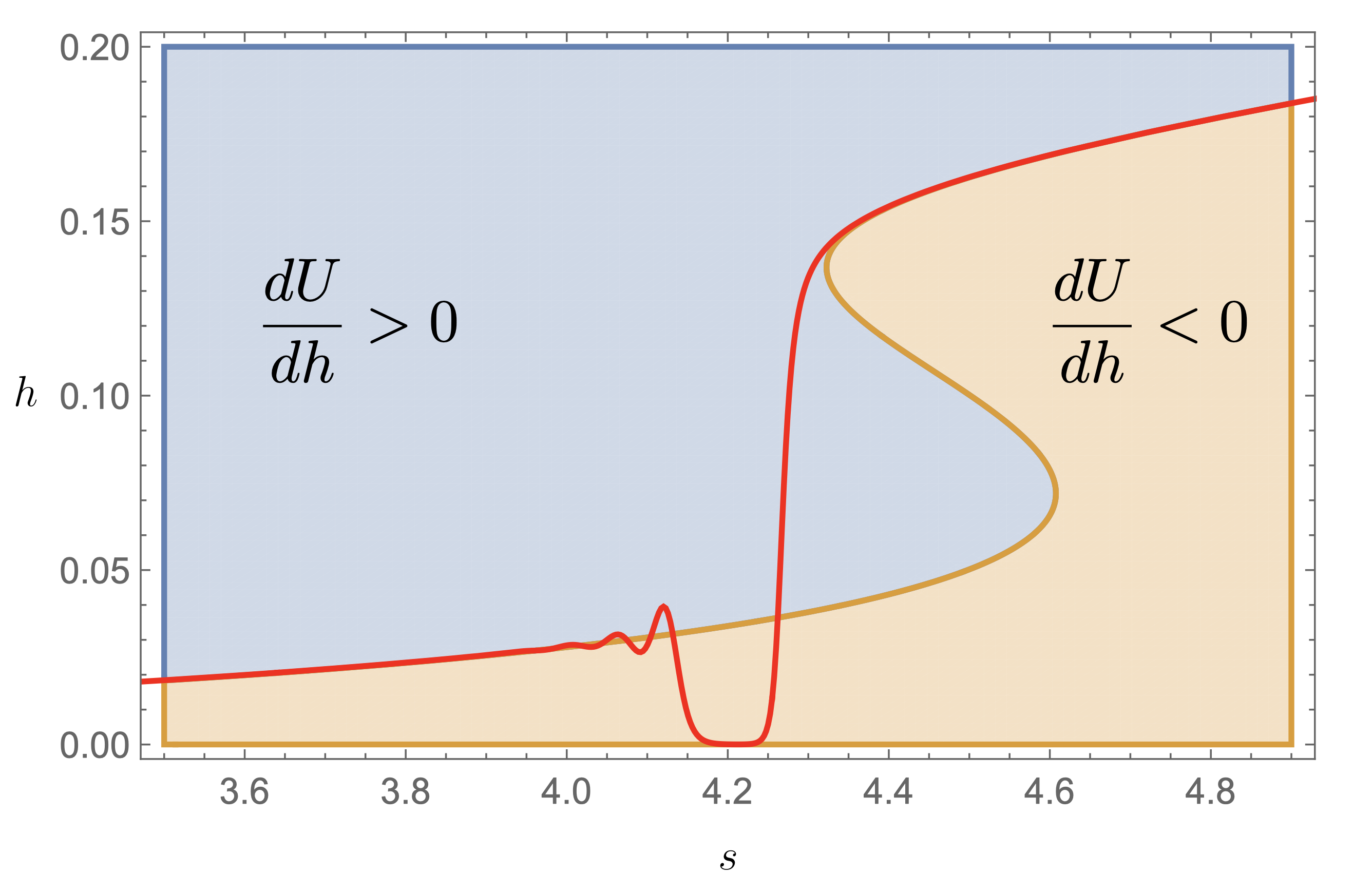}
    \includegraphics[width=.45\textwidth]{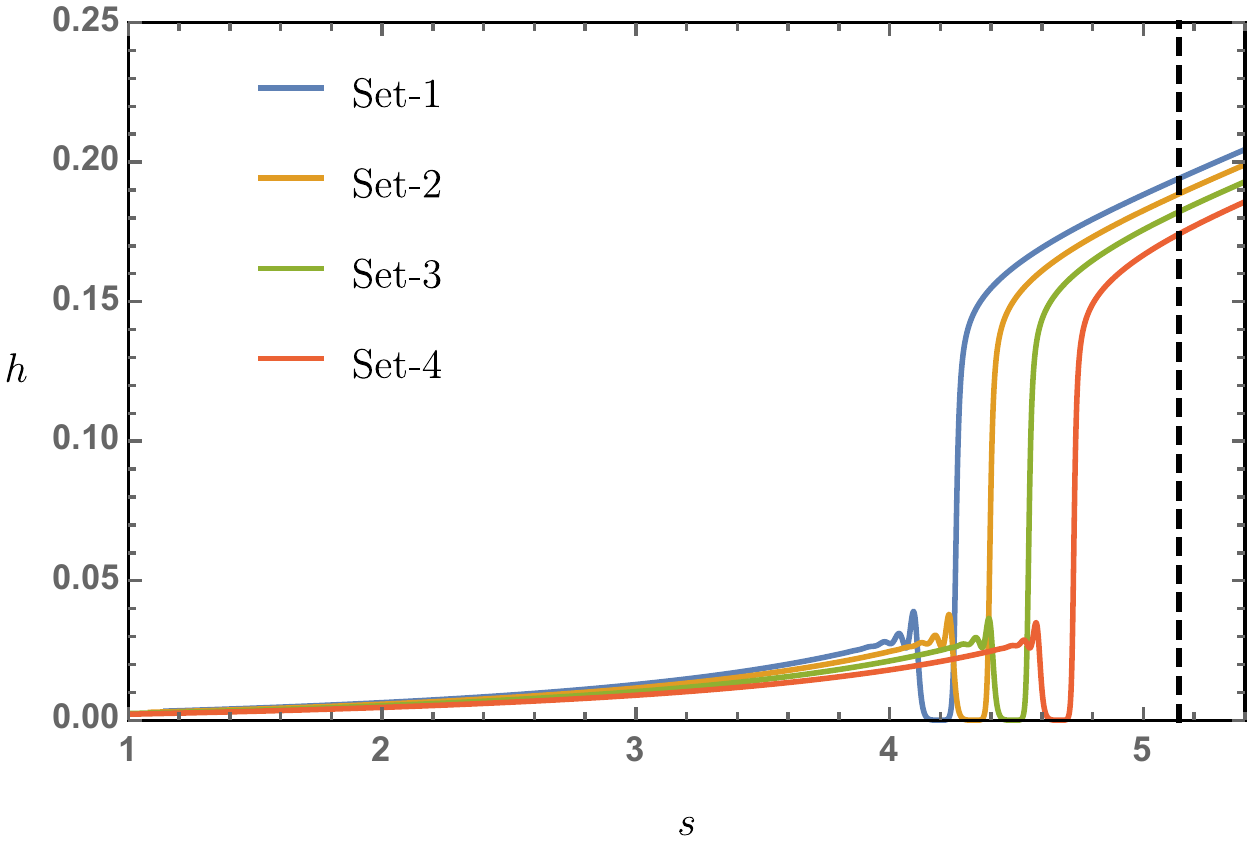} 
   \caption{(Left) Contour of $\frac{dU}{dh} = 0 $, with the trajectory of the inflaton colored in red. Once $h$ reaches $h_{local\,\, min}$, $\frac{dU}{dh}>0$, making the inflaton fall down towards the hill. (Right) Trajectory for several parameter sets, where the dashed black line expresses the field value for $N_\text{inf} = 50$. The position where \textit{hill-climbing} occurs depends on $\xi$ values.  }
   \label{fig:dUdh}
\end{figure}

\subsection{Stage 2.}
This stage contains the initial deviation from the valley. Recall that for the Higgs-$R^2$ potential, the inflaton \textit{initially} follows a valley well defined by $\partial_h U =0$, in which the scalaron field at the valley $s_v$ takes the expression
\begin{equation}
s_{v}  \simeq  \sqrt{\frac{3}{2}} \ln \left[\frac{6 M^2 \xi_0 +2 h^2 \left(\lambda_m+3 M^2 \xi_0 ^2\right) +b\,h^2 \ln \left(\frac{h}{h_m}\right) \left(2 \ln \left(\frac{h}{h_m}\right)+1\right) }{6 M^2 \xi_0 }\right] . 
\end{equation}
This trajectory in general, can have critical points in the $(h,s)$ plane, being
\begin{equation}
\frac{h_{local\,\,max}}{h_m} \simeq e^{-\frac{3}{4} - \frac{\sqrt{5b^2 -16 b \lambda_m -48 b M^2 \xi_0^2 }}{4b}}
 \, \,, \,\, \, \frac{h_{local\,\,min}}{h_m} \simeq e^{-\frac{3}{4} + \frac{\sqrt{5b^2 -16 b \lambda_m -48 b M^2 \xi_0^2 }}{4b}} . 
\label{eq:hmaxhmin}
\end{equation}
Note that these extremal points in the field space exist when the following conditions are satisfied
\begin{equation}
0\leq \xi_0 \lesssim \frac{1}{4\sqrt{3} M} \sqrt{5 b - 16 \lambda_m} \,\, , \,\,\, \lambda_m < \frac{5b}{16} . 
\end{equation}
We focus on the trajectory point $h_{local \,\, min}$. Once the inflaton hits this point, $\partial_h U >0 $ in the order 
\begin{equation}
\left.\frac{\partial U}{\partial h}\right|_{s_v(h_{local\,\,min}),\, h_{local\,\,min}\pm \delta} = A\,\delta^2 + \mathcal{O}(\delta^3) 
\end{equation}
for both $h_{local \,\, min} + \delta$ and $h_{local \,\, min} - \delta$ with $\delta> 0$, with $A$ being a positive constant. Therefore, the inflaton starts rolling down towards the $h$ direction, approaching the hill at $h=0$. This is depicted in figure~\ref{fig:dUdh}. where the figure focuses on the transition region, and it shows that the potential exhibits a region where $dU/dh>0$ giving a roll-down towards the hill.

\subsection{Stage 3. }
\begin{figure}[htbp]
   \centering
   \includegraphics[width=.70\textwidth]{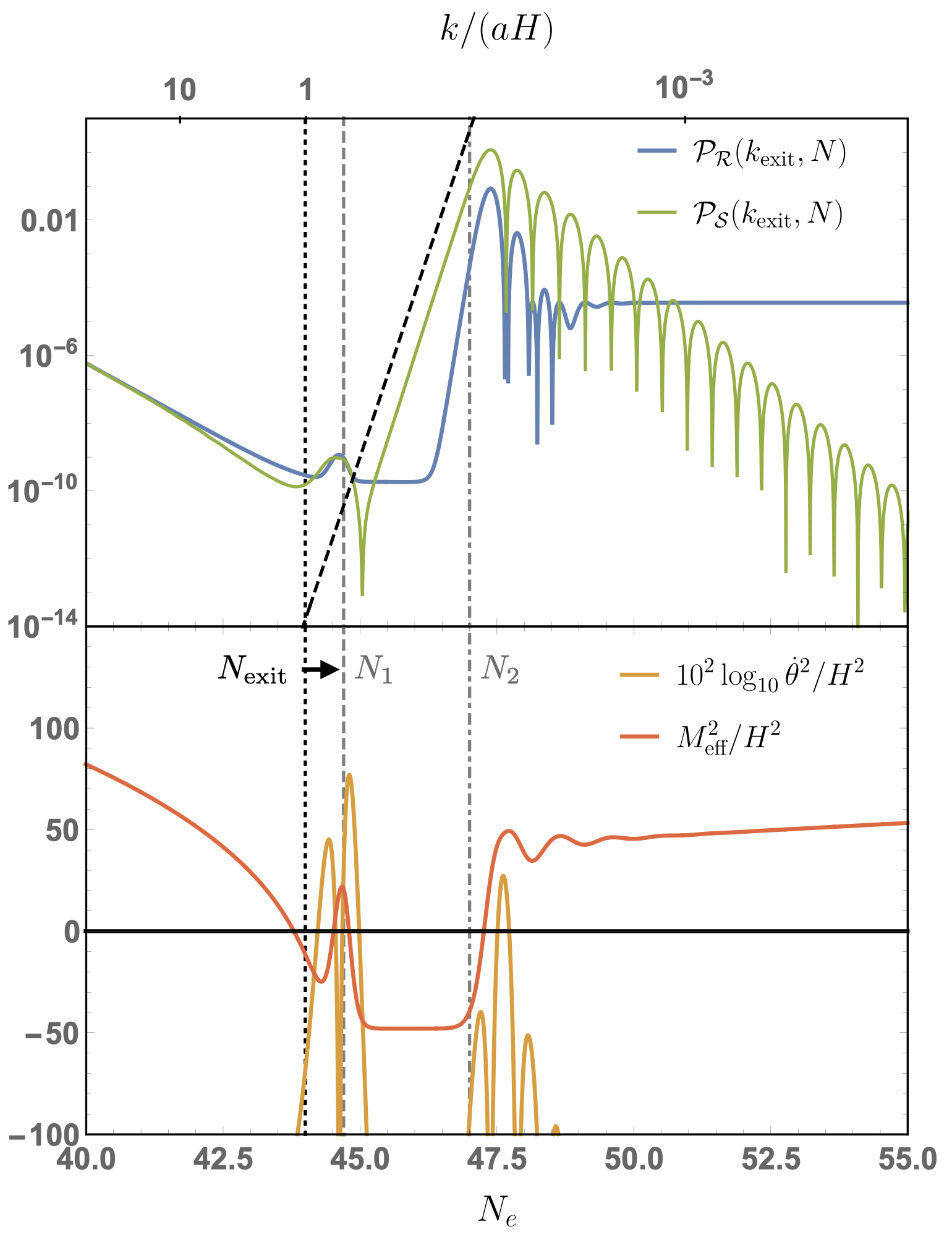}
   \caption{(Top) Perturbation evolution for the $k$ mode that exits the horizon at $N_\text{exit}=44$. The efolds when the inflaton starts and stops rolling down the hill are denoted as $N_1$ and $N_2$ respectively. The black-dashed line represents the growth $\mathcal{P}_\mathcal{S} \propto \exp \left[\left(\frac{2 |M_\text{eff}| }{H}-3\right) N_e \right] $. (Bottom) $M_\text{eff}^2 / H^2 $ and $\dot{\theta}^2 / H^2$ evolution. The tachyonic $M_{\text{eff}}$ induces an exponential growth in the $\mathcal{P}_\mathcal{S}$, and due to the mixing between curvature and isocurvature perturbations this enhancement is translated over to $\mathcal{P}_\mathcal{R}$. }
   \label{fig:perturbation}
\end{figure}

This stage is precisely where the large and negative isocurvature mass induces an exponential growth in the isocurvature perturbation $Q_N$. The evolution of the perturbations are depicted in figure~\ref{fig:perturbation}.

\begin{figure}[htbp]
   \centering
   \includegraphics[width=.80\textwidth]{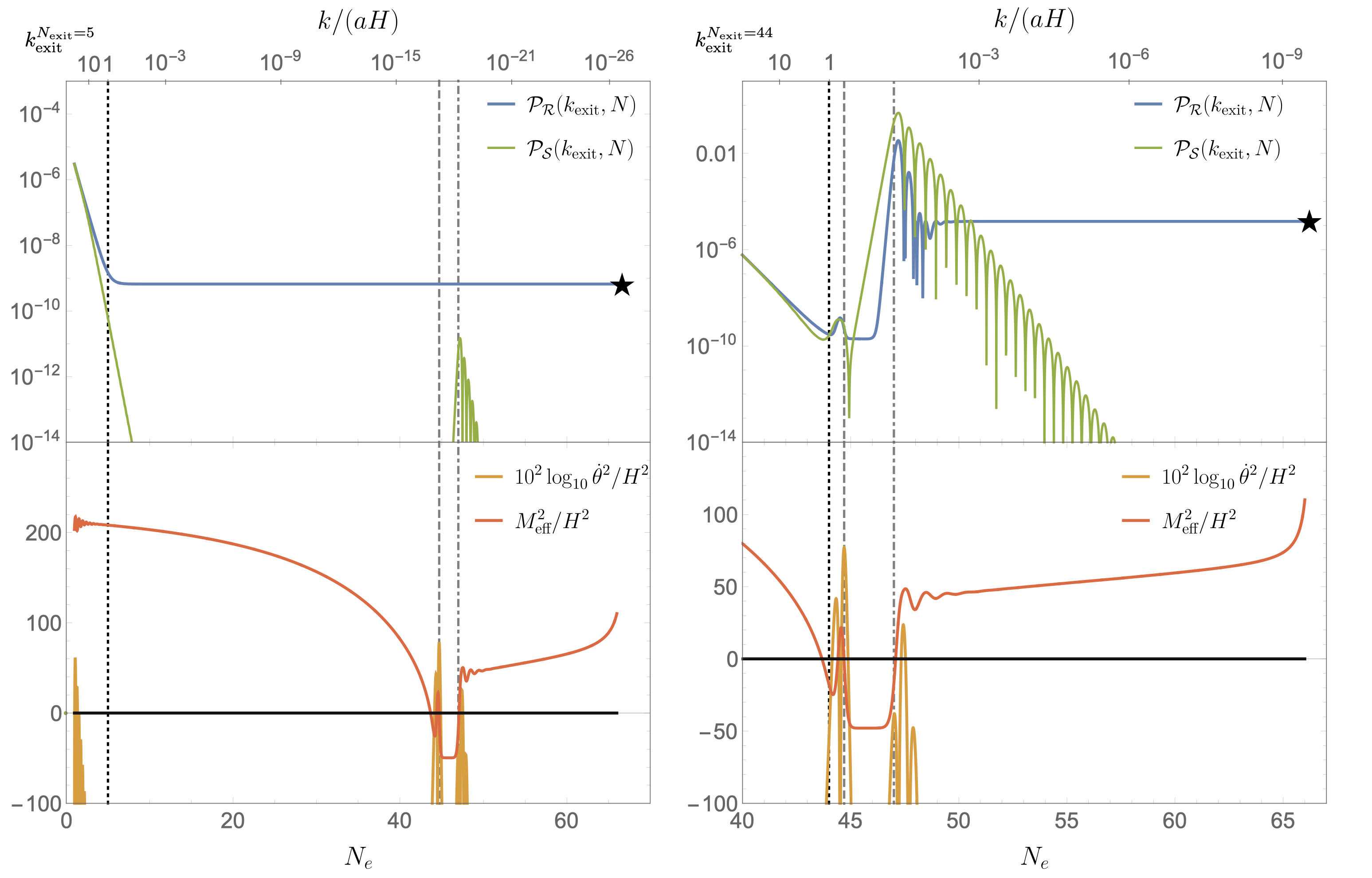}
   \includegraphics[width=.80\textwidth]{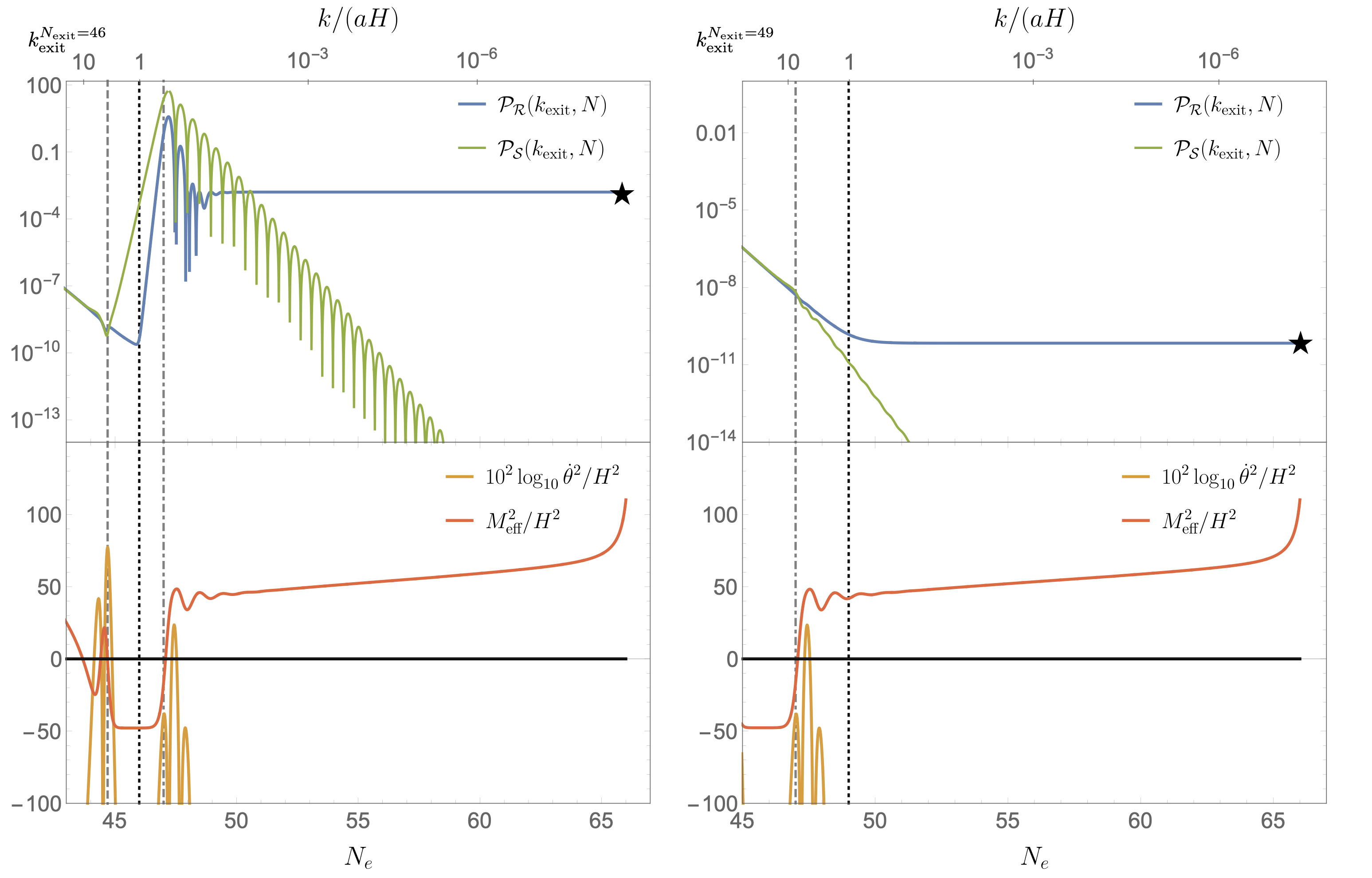}
   \caption{Compilation of the perturbations for $k_\text{exit}$ leaving at certain scales, which corresponds to a definite $N_\text{exit}$. We observe that after the tachyonic increase of the isocurvature, and consequentially the curvature perturbation, the isocurvature part exponentially decays, and the perturbations become \textit{adiabatic} at scales evaluated at the end of inflation ($N_\text{end} $ in the plot, shown at the star).   }
   \label{fig:perturbationcollection}
\end{figure}

From the first turn, due to the mixing term in the equations of motion, the perturbations mix and as $\dot{\theta}^2 / H^2 >1$, the $\mathcal{R}$ experiences a slight bump in its evolution, however its effect is negligible. 

The period when the inflaton rolls down the hill, where the starting efolds at $N_1$ and the end efolds at $N_2$, the isocurvature mass takes $M_\text{eff}^2 / H^2 \ll 0$, therefore the isocurvature perturbation $Q_N$ is dominated by the exponential growth from the negative isocurvature mass. Recalling the isocurvature perturbations equations eq.~(\ref{eq:isocurvaturepert}), eq.~(\ref{eq:isocurvaturepertstage1}) while neglecting the source terms ,as this is precisely the case when the inflaton rolls down the $h=0$ \textit{hill}, 
\begin{equation}
\frac{d^2 {Q}_{N,k}}{dN_e^2} + 3\frac{d {Q}_{N,k}}{dN_e} + \left(\frac{k^2}{a^2 H^2} - \frac{|M_\text{eff}^2|}{H^2} \right)Q_{N,k} = 0 
\end{equation}
with the solutions
\begin{equation}
{Q}_{N, k} (N_e) = e^{-\frac{3}{2} N_e}\left[d_3\,e^{-\frac{N_e}{2}  \sqrt{9  -4 \frac{M_\text{eff}^2}{H^2} -4 \epsilon_k^2 } } + d_4 \, e^{\frac{N_e}{2}  \sqrt{9  -4 \frac{M_\text{eff}^2}{H^2} -4 \epsilon_k^2  } }  \right] \xrightarrow[|M_\text{eff}^2| \gg H^2]{\epsilon_k^2 \ll 1 }d_4 ~ e^{\left(\frac{|M_\text{eff}|}{H} - \frac{3}{2}\right) N_e }
\end{equation}
with the $M_\text{eff}^2$ in this period taking the form of eq.~(\ref{eq:meffapprox}). Consequentially, 

\begin{align}
\mathcal{P}_{\mathcal{S}}(k_\text{exit}, N_e) = \frac{k_\text{exit}^3}{2\pi^2} \frac{H^2}{\dot{\phi}_0^2} \langle Q_{N, k} \,Q_{N, k}\rangle= \mathcal{P}_{\mathcal{S}}(k_\text{exit}, N_1) ~e^{\left(\frac{2|M_\text{eff}|}{H} - {3}\right)  (N_e - N_1) }
\end{align}
therefore  $Q_{N,k}$, and consequentially $\mathcal{P}_\mathcal{S}$  grows \textit{exponentially} during this period. $\mathcal{R}$, however, does not grow instantaneously, precisely due to the fact that $\dot{\theta}^2 /H^2 \ll 1$ in this period. Then, as the second turn with $\dot{\theta}^2 /H^2 \gg 1$ occurs, the enhanced $Q_N$ is then sourced to $\mathcal{R}$, now also exponentially growing and decaying according to the mixing, will then stop evolving in the superhorizon limit when  $\dot{\theta}^2 /H^2 \ll 1$ occurs again. 

One may raise the following question:\textit{is eq.~(\ref{eq:lambdarunning}) a sufficient approximation for this scenario, judging by the fact that $h$ approaches 0.} The answer is yes. The turn in the trajectory  occurs around eq.~(\ref{eq:hmaxhmin}) where $h_{local \,\, min} \sim h_m$. The region where $h \ll h_m$, the potential term $U \supset \frac{\lambda(h)}{4} h^4 e^{-2\sqrt{\frac{2}{3}} s}$ is subdominant. 

\section{Power spectrum and PBH abundance}
We numerically compute the cosmological perturbations in this scenario using PyTransport \cite{Mulryne:2016mzv}. Here we present several parameter sets exhibiting this large local feature in the power spectrum, enlisted in Table \ref{tab:i}. The Planck CMB pivot scale is set as $k_* = 0.05~\text{Mpc}^{-1}$.

\begin{figure}[tbp]
   \centering
   \includegraphics[width=0.95\textwidth]{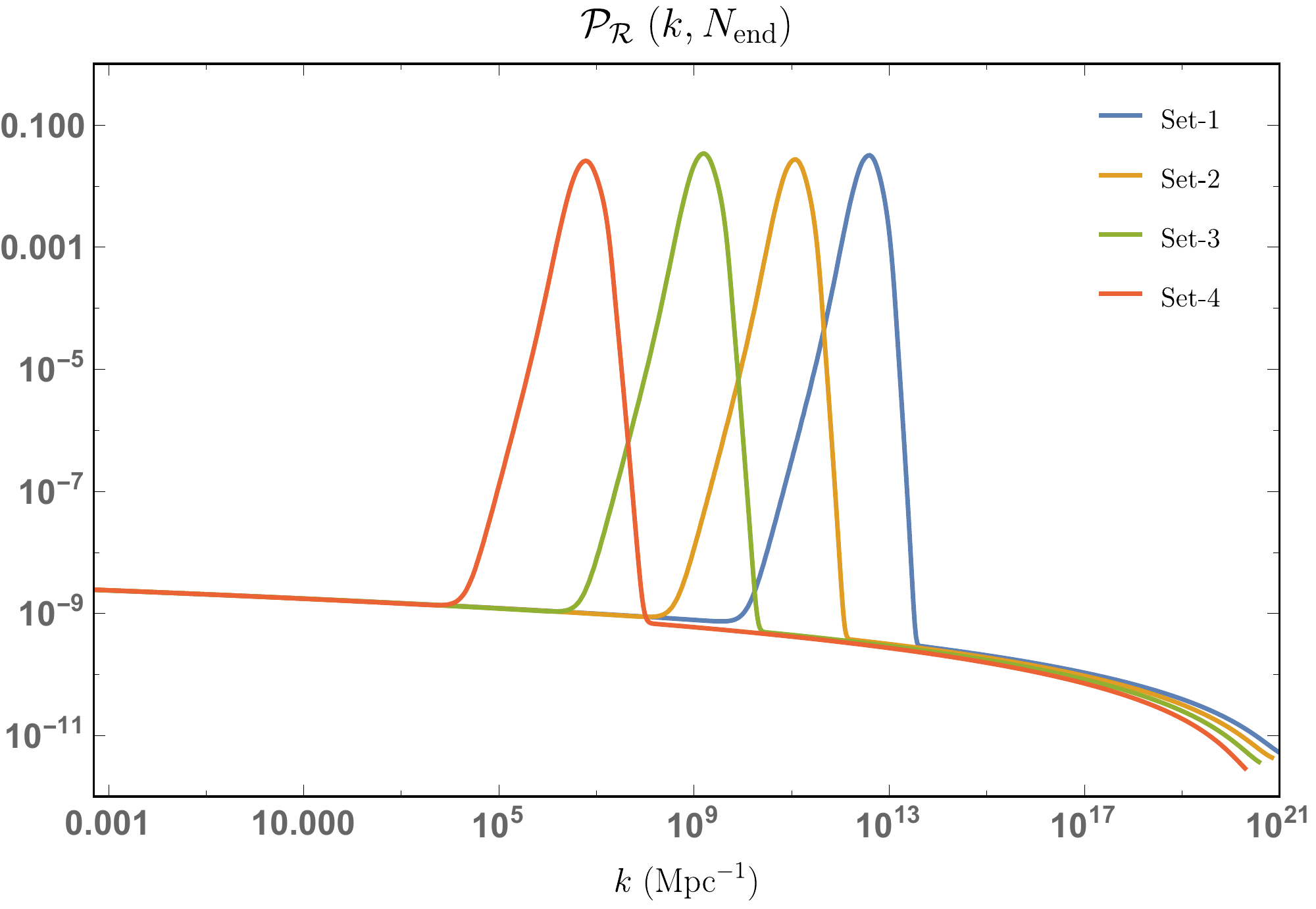} 
   \caption{The observational curvature power spectrum  $\mathcal{P}_\mathcal{R} (k, N_\text{end})$ for the benchmark parameter sets in Table \ref{tab:i}.  }
   \label{fig:powerspectrum}
\end{figure}

\begin{table}[htbp]
\centering
\resizebox{1\textwidth}{!}{
\begin{tabular}{|c |c c c c c c | c c c c |}
\hline
$ \text{Set}$ &  $M (M_P) $ &$\xi_0$ &$\lambda_m (\times 10^{-6})$ & $\beta_2 $ & $\beta_\xi^0$ & $h_m\,(M_P)$  &$k_{max} \,(\text{Mpc}^{-1})$  &$ \mathcal{P}_{\mathcal{R}, max} $& $n_s$ & $r$ \\
\hline
$1$ & $1.3 \times 10^{-5} $ &$4.0$ &$4.1743336$ & $0.5$ & $ -0.01$& $0.21$ & $3.9\times 10^{12}$ &$0.032$ & $0.967$ & $0.004$ \\ 
$2$ & $1.3 \times 10^{-5} $ &$3.5$ &$ 4.1003376$ & $0.5 $ & $ -0.01$& $0.21$ & $1.2\times 10^{11} $ &$0.027$ & $0.967$ & $0.004$\\ 
$3$ & $1.3 \times 10^{-5} $ &$3.0$ &$ 4.0109148$ & $0.5$ & $ -0.01$& $0.21$ & $1.6\times 10^{9} $&$0.034$ & $0.967$ & $0.004$ \\ 
$4 $ & $1.3 \times 10^{-5} $ &$2.5$ &$3.8998765$ & $0.5$ & $ -0.01$& $0.21$ & $6.5 \times 10^{6} $&$0.02$ & $0.967$ & $0.004$ \\ 
\hline
\end{tabular}
}
\caption{\label{tab:i} Several benchmark parameter sets ($M$, $\xi_0$, $\lambda_m$, $\beta_2 = (4\pi)^4\,b$, $\beta_\xi^0 = b_\xi / 2$,  $h_m$) and their corresponding small-scale observables. }
\end{table}

\begin{figure}[h!]
   \centering
   \includegraphics[width=0.87\textwidth]{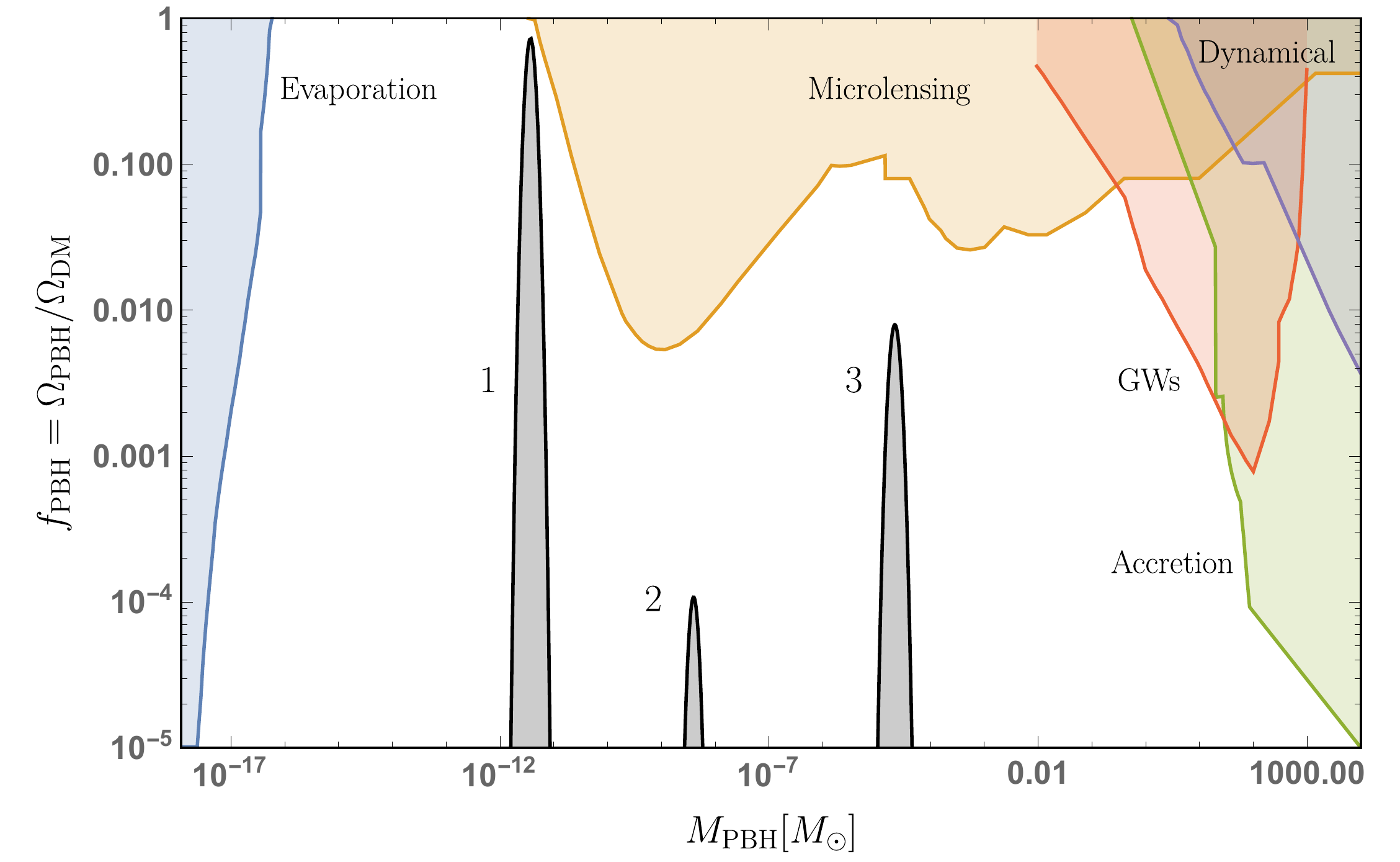} 
   \caption{{The corresponding PBH abundances for the benchmark parameter sets in Table \ref{tab:i}. The observational constraints are taken from~\cite{Carr:2020gox, bradley_j_kavanagh_2019_3538999}. Depending on the parameters the tachyonic instability-induced perturbations one can induce scenarios where PBHs can consist a significant amount of dark matter within the constraint bounds. Set-4 is not depicted in the figure range due to a smaller peak value $\mathcal{P}_{\mathcal{R}, max}$.  }}
   \label{fig:PBHabundance}
\end{figure}

The corresponding curvature power spectrum $\mathcal{P}_\mathcal{R}(k)$ for these parameter sets are depicted in figure~\ref{fig:powerspectrum}. Each exhibit a near-scale invariant power spectrum with $n_s$ giving perfect consistency with current Planck CMB observations~\cite{Planck:2018vyg, Planck:2018jri}. Note that the CMB predictions also are consistent among parameters, regardless of the position of the localized peak. 
This is precisely due to the fact that the \textit{tachyonic} enhancement of the curvature/isocurvature perturbations induces an exponential increase, requiring a much shorter period on how long this instability sustains.
This amplified $\mathcal{P}_\mathcal{R}$ at small scales can also lead to copious PBH production, which depending on the mass of the PBH can account for the majority of the dark matter in our universe. 
{Taking a peaks theory approach~\cite{Bardeen:1985tr, Green:2004wb, Young:2014ana, Yoo:2018kvb, Yoo:2019pma, Young:2020xmk, Yoo:2020dkz, Wang:2021kbh}, where the details are in Appendix~\ref{appendix:PBH_abundance}.,  we depict the PBH abundance {$f_{\text{PBH}} ( M_{\text{PBH}})  = \Omega_{\text{PBH}}/\Omega_\text{DM}$} in figure~\ref{fig:PBHabundance}., taking a common critical density contrast $\delta_c = 0.41$ \cite{Harada:2013epa, Harada:2015yda} and a Gaussian window function $W(k,R) = \exp \left(- \frac{k^2 R^2}{2} \right)$. One major feature the tachyonic instability-induced $\mathcal{P}_\mathcal{R}$ enhancement composes is that it can cover all viable mass ranges of PBHs, which is in contrast with our previous USR induced PBH studies \cite{Cheong:2019vzl}.}

 \begin{figure}[tbp]
	\centering
	\includegraphics[width=0.6\textwidth]{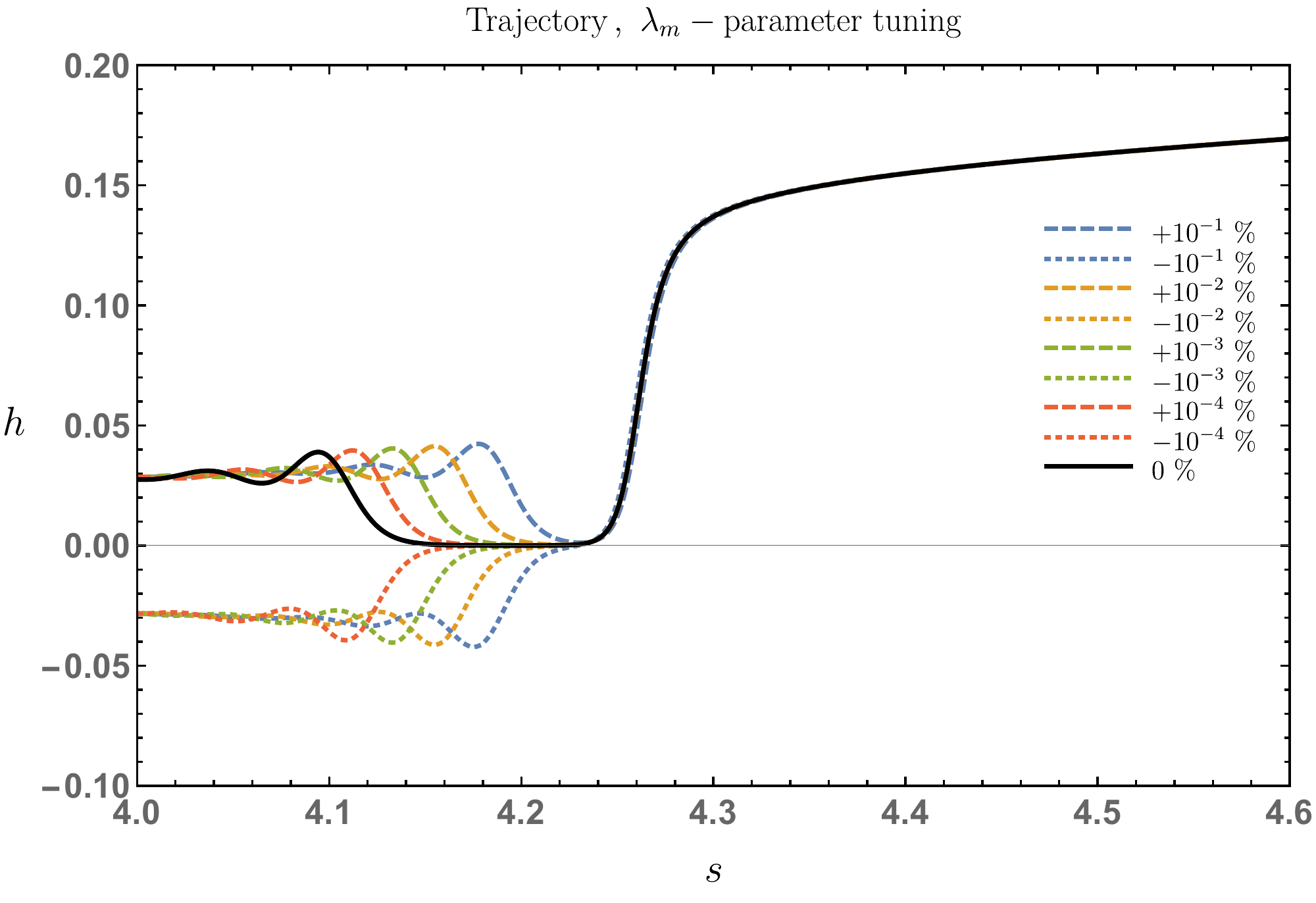}
	\caption{Parameter sensitivity on the inflaton trajectory. For demonstration, we take the parameter $\lambda_m$ tuning for Set. 1 in Table \ref{tab:i}. The percentage level corresponds to the parameter $\delta \lambda_m / \lambda_m$. }
	\label{fig:trajectory_tuning}
\end{figure}

 \begin{figure}[tbp]
	\centering
\includegraphics[width=0.45\textwidth]{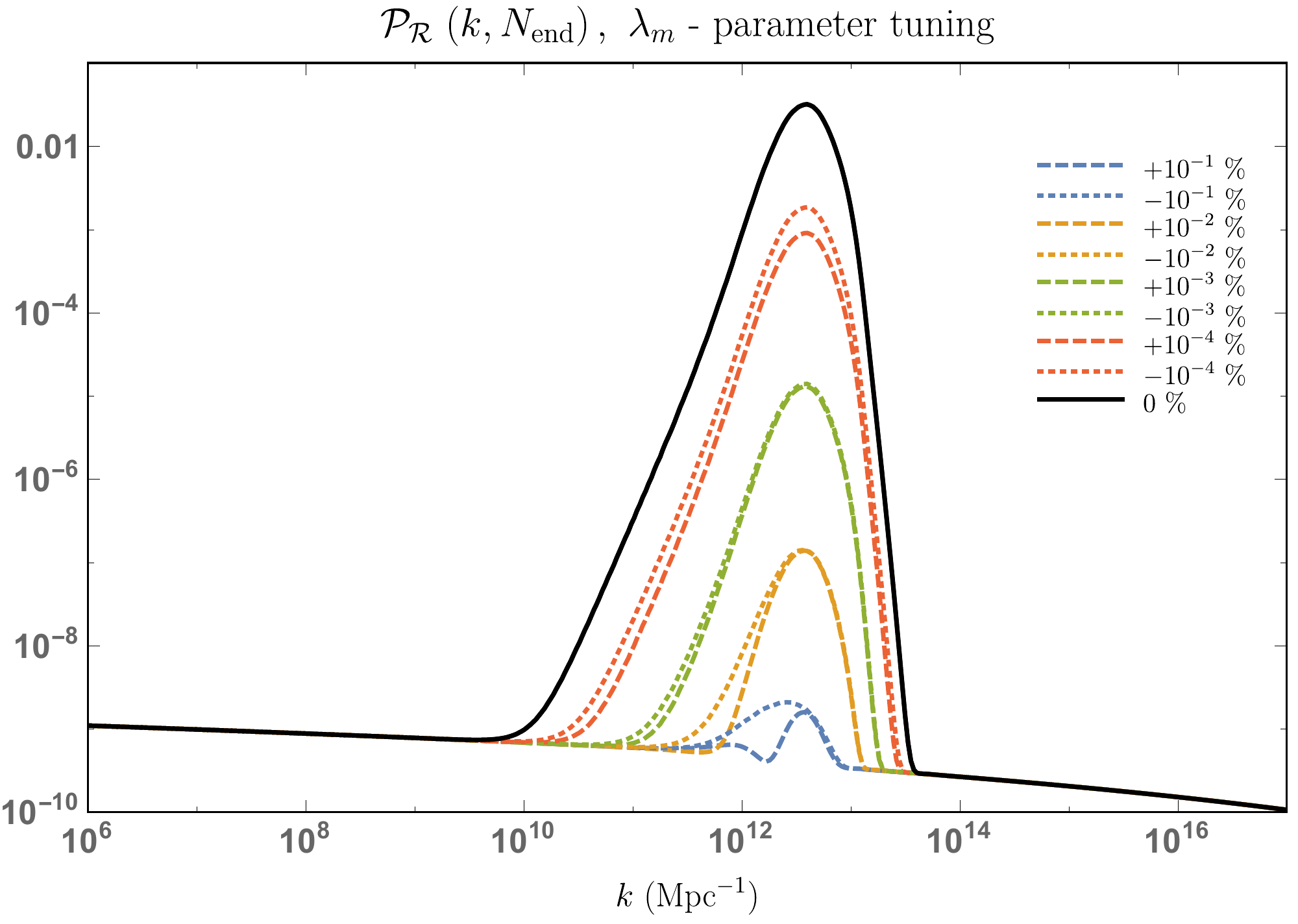} 
\includegraphics[width=0.45\textwidth]{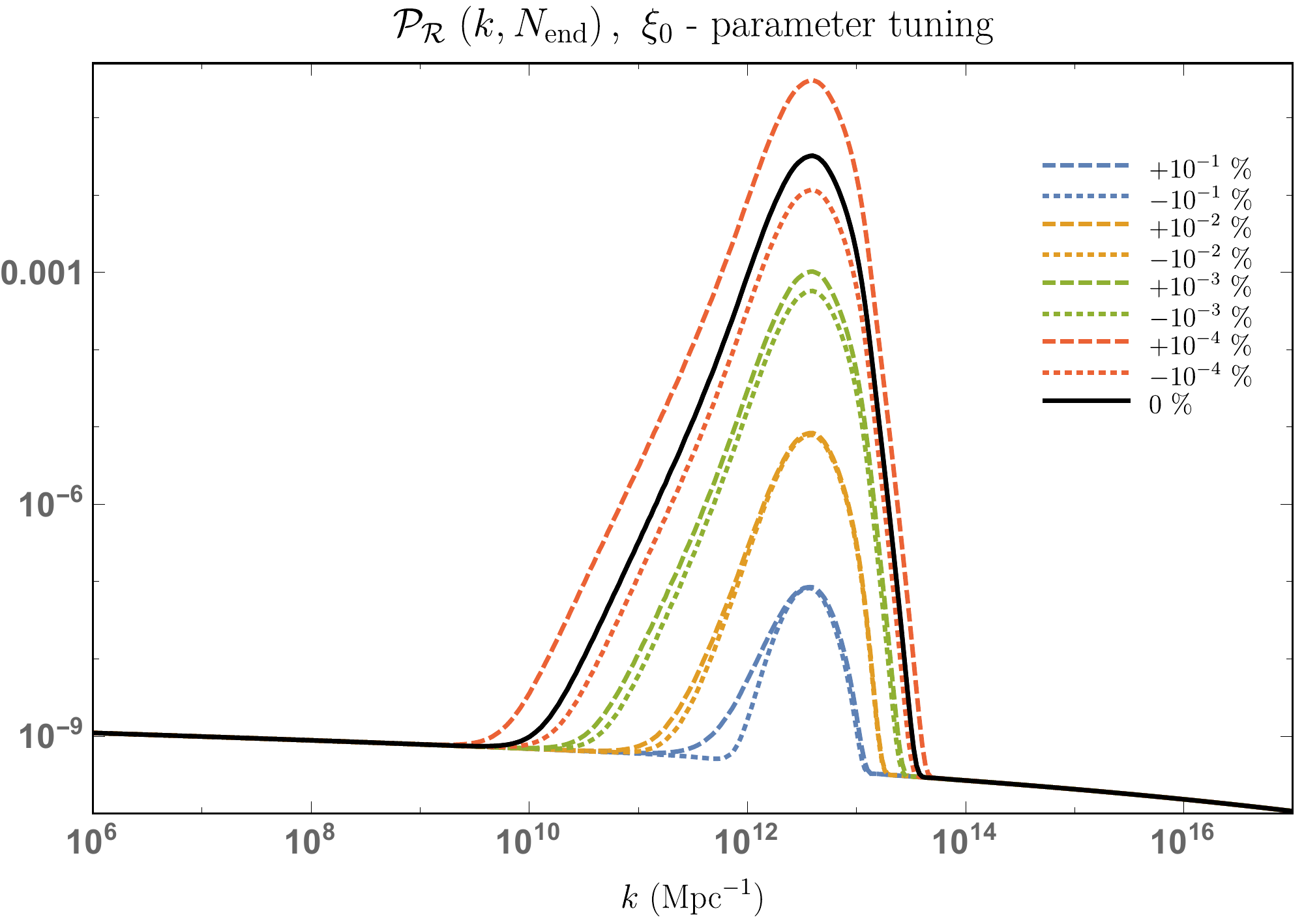} 
	\caption{(Left )The degree of parameter $\lambda_m$ tuning for Set. 1 in Table \ref{tab:i}. The percentage level corresponds to the parameter $\delta \lambda_m / \lambda_m$.  (Right) The degree of parameter $\xi_0$ tuning for Set. 1 in Table \ref{tab:i}. The percentage level corresponds to the parameter $\delta \xi_0 / \xi_0$. }
	\label{fig:powerspectrum_tuning}
\end{figure}

\subsection{Degree of parameter tuning}  
Obviously, the tachyonic enhancement is subject to parameter tuning. The evolution of the inflaton along the hill determines the amount of enhancement, and as this period itself is unstable the parameters will need some degree of tuning to have a noticeable perturbation growth.

{To show this, we choose the Set 1.  parameters and vary $\lambda_m$ and $\xi_0$ and explicitly allowing the $h<0$ region.\footnote{We take $\lambda_m$ and $\xi_0$  as the tuning parameter as it encapsulates the role on sustaining the instability period. } As depicted in figure~\ref{fig:trajectory_tuning}., the field evolution along the $h=0$ hill is sensitive to the parameter choices. This difference to the evolution period on the hill leads to a difference in the curvature power spectrum peak, depicted in figure~\ref{fig:powerspectrum_tuning}. Notice that in order to sustain an enhancement in the curvature power spectrum to be $\mathcal{P}_{\mathcal{R}} \sim 10^{-2}$, the parameter $\delta \lambda_m / \lambda_m \equiv  (\lambda_m^{dev} - \lambda_m)/\lambda_m \sim  \mathcal{O}(10^{-4}) \,\%$, with $\lambda_m^{dev}$ being the parameter that deviates from the benchmark set parameter, which is in similar orders with fine-tuning degrees in single-field polynomial inflation models as well \cite{Hertzberg:2017dkh}. The $\xi_0$ parameter, on the other hand, is an order less sensitive to achieve the same order of enhancement, with $\delta \xi_0 / \xi_0 \equiv  (\xi_0^{dev} - \xi_0)/\xi_0 \sim  \mathcal{O}(10^{-3}) \,\%$, Noticeably, if one requires the detectability of the SGWB, the required enhancement softens to $\mathcal{P}_{\mathcal{R}} \sim 10^{-4} \, (10^{-5})$, hence the tuning of the parameters also decrease to an order less to be $\delta \lambda_m / \lambda_m \sim \mathcal{O}(10^{-3})\, \%$,  $\delta \xi_0 / \xi_0 \sim \mathcal{O}(10^{-2})\, \%$. }

\section{Stochastic gravitational wave background (SGWB) at the second order}

%
 \begin{figure}[tbp]
	\centering
	\includegraphics[width=0.9\textwidth]{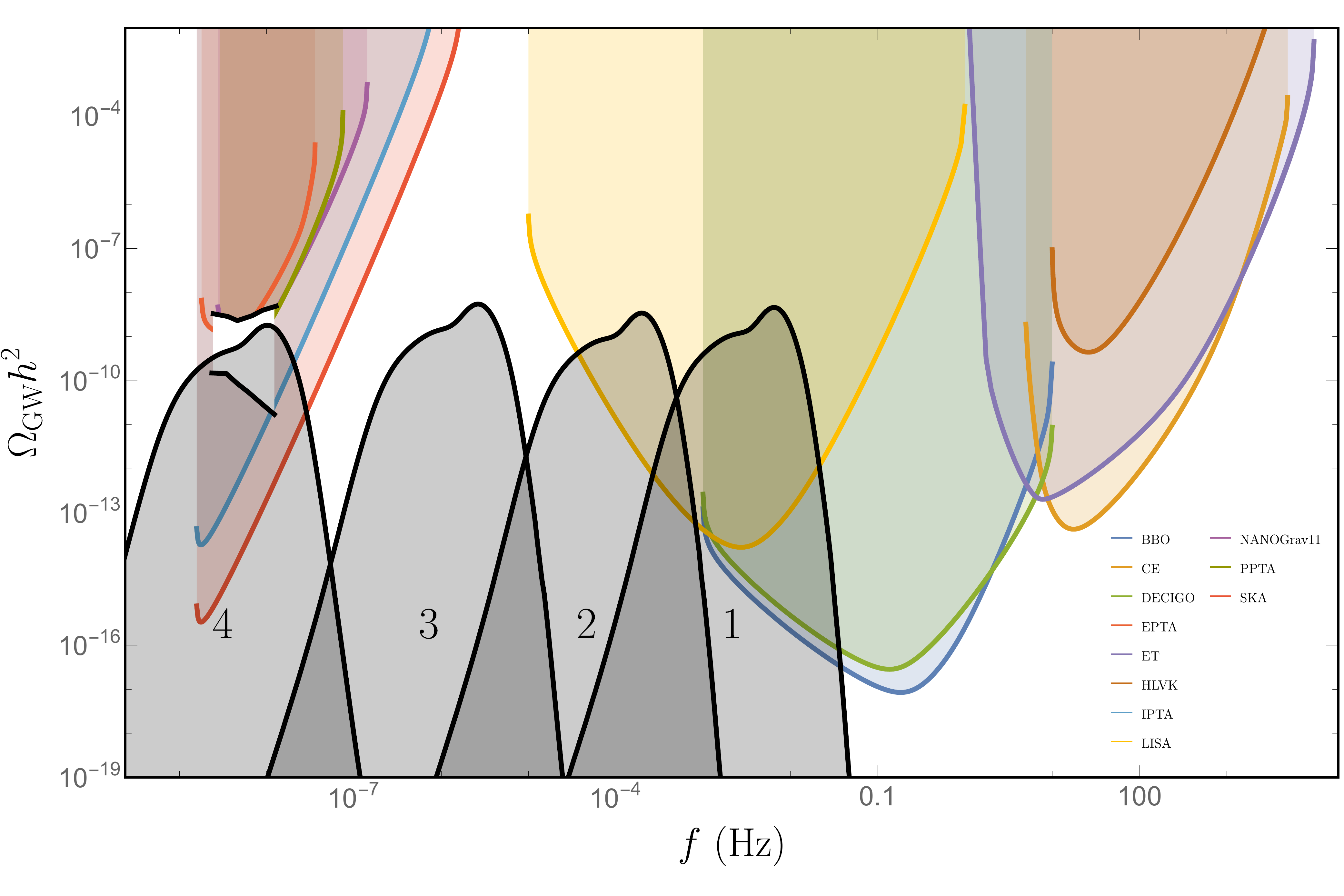} 
	\caption{The corresponding GW abundances for the benchmark parameter sets in Table \ref{tab:i}, overlayed over various gravitational wave observatory's sensitivity curves~\cite{Schmitz:2020syl, schmitz_kai_2020_3689582}. The white region at nano-Hertz frequency corresponds to the NANOGrav 12.5 year results~\cite{NANOGrav:2020bcs}.}
	\label{fig:GWspectrum}
\end{figure}
%

The amplified curvature perturbations also lead to a copious amount of stochastic GWs. The current energy density fraction per logarithmic wavelength of the GWs from second order is, following~\cite{Espinosa:2018eve, Kohri:2018awv}
\begin{align}
\Omega_\text{GW}(\eta_0 , k) = c_g \frac{\Omega_{r,0}}{6} \int_{0}^{\infty} dv \int_{|1-v|}^{1+v} du \left(\frac{4v^2 - (1+v^2 - u^2)^2}{4uv} \right)^2 \overline{\mathcal{I}^2 (v,u)} \mathcal{P}_{\mathcal{R}} (kv) \mathcal{P}_\mathcal{R} (ku )
\end{align}
where {$\Omega_{r,0}  \approx 5.38 \times 10^{-5} $ is the current radiation energy density fraction~\cite{Planck:2018vyg}, $x \equiv  k \eta$ with $\eta$ being the conformal time,  $c_g \equiv \frac{a_f^4 \rho_r (\eta_f)}{\rho_r (\eta_0)} = \frac{g_*}{g_*^0 } \left(\frac{{g_*^0}_S}{{g_*}_S}\right)^{4/3} \approx 0.4 $ by taking the current universe effective energy and entropy degree of freedom as $g_*^0 = 3.36$ and ${g_*^0}_S = 3.91$ respectively. The degree of freedom at the evaluation of perturbations take the value ${g_*} = {g_*}_S = 106.75 $. $\overline{\mathcal{I}^2 (v, u )}$ is expressed analytically in radiation-domination as~\cite{Espinosa:2018eve, Kohri:2018awv, Inomata:2019yww}. 
\begin{align}
\overline{\mathcal{I}^2 (v, u)} = \frac{1}{2} \left[\frac{3(u^2 + v^2  - 3 )}{4 u^3 v^3} \right]^2 & \left[  \left(-4 u v + (u^2 + v^2 -3) \ln \left|\frac{3 - (u+v)^2 }{3- (u-v)^2 } \right| \right)^2  \right. \nonumber \\ & + \left. \pi^2 (u^2 +v^2 -3)^2 \Theta (v+u - \sqrt{3}) \right] . 
\end{align}}
The current universe SGWB energy fraction $\Omega_\text{GW} \,h^2$ for our benchmark parameters are depicted in figure~\ref{fig:GWspectrum}. Note that due to the nature of the tachyonic instability, the GW spectrum can span over all frequencies from aLIGO frequencies all the way up to PTA frequencies. Utilizing current and future GW observatories, we can obtain useful information on the $\xi$ and $\lambda(h)$ parameters, ultimately gaining information on the running behavior of the SM Higgs self coupling for high energy scales. Noticeably, our Set. 4 parameters directly correspond to the recently reported NANOGrav results. Therefore, these running parameters and nonminimal coupling values, if the results are confirmed to be indeed a SGWB, will be able to be a possible source of the observed SGWB.


\section{Conclusion and Discussions}
In this work we demonstrated that the $\lambda$ running in Higgs-$R^2$ inflation can be implemented in the effective production of enhanced curvature perturbations, consequentially leading to amplified second order stochastic gravitational wave productions and possibly primordial black holes. The two field potential characterized by the scalaron and Higgs $(s,h)$ exhibits a temporary valley structure breakdown induced by the running of $\lambda(h)$. The inflaton then rolls down towards the hill at $h=0$, where there is a tachyonic instability that depends on the scalaron mass $M$ and the nonminimal coupling $\xi$. Isocurvature perturbations are exponentially enhanced, which are transferred to the curvature perturbation as the inflaton rolls back down the hill and settles at the lower field valley. Compared to our previous work that implemented a ultra-slow-roll phase \cite{Cheong:2019vzl}, the tachyonic enhancement presented in this work occurs in a much shorter duration, hence allowing for effectively all mass ranges of primordial black hole dark matter without conflicting Planck CMB observables. It also allows parameters that coincide with the recently reported stochastic process by NANOGrav and other PTA observatories.

There are several aspects to foresee from here. In this work we effectively parameterized the $\lambda$ running with the parameters $\lambda_m,~ b,~ h_m$. In principle these will correspond to the physical running parameters $m_{top},~\alpha_s,~m_W$ at low energies, therefore providing a connection between the low energy SM parameters and the inflationary features. A full parameter scan of these SM parameters that incorporate a sizable $\mathcal{P}_\mathcal{R}$ at a certain scale will provide information on the running parameters of our Standard Model Higgs, which we seek to pursue in a future work~\cite{Cheong:20xxabc}. 

Also, in this work we only considered Gaussian fluctuations in the cosmological perturbations. It is well known that multi-field inflation, especially those incorporating a tachyonic instability may exhibit high levels on non-Gaussianity, which will alter the predictions of SGWB, and PBH abundances~\cite{Renaux-Petel:2015mga, Garcia-Saenz:2018ifx, Garcia-Saenz:2018vqf, Fumagalli:2019noh, Renaux-Petel:2021yxh}. We leave the computation of non-Gaussianity and its impact on small scale observables for a future work.

\acknowledgments

We thank Minxi He, Sung Mook Lee, Shi Pi, Misao Sasaki for precious discussions. This work was supported by JSPS KAKENHI grant number JP17H01131 (K.K.), MEXT KAKENHI grant
number JP20H04750 (K.K.), and the National Research Foundation grants funded by the Korean government (MSIP) (NRF-2019R1A2C1089334)\&(NRF-2021R1A4A2001897) (SCP).

\appendix
\section{Peaks theory approach on the PBH abundance calculation. }
\label{appendix:PBH_abundance}

In this appendix we address the peaks theory approach to calculate the PBH abundance. The density contrast power spectrum is related to the curvature power spectrum through the following formula
\begin{align}
\mathcal{P}_{\delta} (k ) = \frac{4(1+w)^2}{(5+3w)^2 } \left( \frac{k}{aH}\right)^4 \mathcal{P}_{\mathcal{R}}(k) \rightarrow \frac{16}{81} \left( \frac{k}{aH}\right)^4 \mathcal{P}_{\mathcal{R}}(k) \,\,\, \text{ in RD : $  w = \frac{1}{3} $ . } 
\end{align}
We then smooth the density contrast to a typical scale $R= \frac{1}{aH}$, with a window function $W(k, R)$ in Fourier space. Hence, the spectral index of the density contrast with a window function will take the form
\begin{align}
\sigma_i^2 (R ) = \int_0^{\infty} \frac{dk}{k} k^{2i} W^2(k,R) \mathcal{P}_\mathcal{\delta}  = \frac{16}{81} \int_0^{\infty} \frac{dk}{k} k^{2i} W^2 (k,R)\mathcal{P}_\mathcal{R}
\end{align}
with the $i=0$-th moment corresponding to the variance $\sigma_0^2 (R) \equiv \langle \delta^2 (\mathbf{x} , R) \rangle$ . 
The PBH mass can be expressed as a function to the associated horizon mass 
\begin{align}
M_\text{PBH} = \mathcal{K} M_H = \frac{\mathcal{K}}{2 G H}
\end{align}
with $ M_H $ being the horizon mass, $\mathcal{K}$ being the collapse efficiency. We take $\mathcal{K} = 0.2$. Associating the mass with the relevant wave number, we get the following scaling 
\begin{align}
M_\text{PBH} = 4.64\times 10^{15}~{\gamma} \left(\frac{g_*}{106.75}\right)^{-\frac{1}{6}} \left( \frac{k_\text{PBH}}{k_*}\right)^{-2} M_{\odot}
\end{align} 
with $ k_* = 0.05~\text{Mpc}^{-1}$ denoting the CMB pivot scale and $g_*= 106.75$ the relativistic degrees of freedom at that epoch. 

In peaks theory~\cite{Bardeen:1985tr} ( see also~\cite{Green:2004wb, Young:2014ana, Yoo:2018kvb, Yoo:2019pma, Young:2020xmk, Yoo:2020dkz, Wang:2021kbh}) , the PBH mass fraction $\beta_\text{PBH} (M) \equiv \left. \frac{\rho_\text{PBH} (M) }{\rho_\text{tot}} \right|_{\text{formation}}$ is related to the peak number density over a criteria $\nu > \nu_c$ 
\begin{align}
n( \nu_c )  = \frac{1}{(2 \pi )^2 } \left( \frac{\sigma_2}{\sqrt{3} \sigma_1}\right)^3 \int_{\nu_c}^{\infty} d\nu \int_0^{\infty} d\xi_1 \frac{f(\xi_1)}{\sqrt{2\pi (1 - \gamma^2)}} \exp{\left[ - \frac{1}{2}\left(\nu^2 + \frac{(\xi_1^2 - \gamma \nu)^2}{1 - \gamma^2} \right) \right]} 
\end{align}
leading to the relation 
\begin{align}
\beta_\text{PBH}( R ) &= n (\nu_c) ( 2 \pi)^{3/2} R^3  \nonumber \\ 
&= \frac{1}{\sqrt{2\pi}} \left(\frac{R \sigma_2}{\sqrt{3} \sigma_1} \right)^3 \int_{\nu_c}^{\infty} d\nu \int_0^{\infty} d\xi_1 \frac{f(\xi_1)}{\sqrt{2\pi (1 - \gamma^2)}} \exp{\left[ - \frac{1}{2}\left(\nu^2 + \frac{(\xi_1^2 - \gamma \nu)^2}{1 - \gamma^2} \right) \right]}. 
\end{align}
The parameters stated here are $\nu = \frac{\delta}{\sigma_0 }$, $\nu_c = \frac{\delta_c}{\sigma_0 }$,$\gamma = \frac{\sigma_1^2 }{\sigma_0 \sigma_2 }$, and the $f(\xi_1) $ function being 
\begin{align}
f(\xi_1) = \frac{1}{2 }(\xi_1^3 -& 3\xi_1) \left(\text{erf} \left[\sqrt{\frac{5}{2}}\xi_1  \right] + \text{erf} \left[ \sqrt{\frac{5}{8}} \xi_1 \right]\right) \nonumber \\
&+ \sqrt{\frac{2 }{5\pi}} \left\{\left( \frac{8}{5} + \frac{31}{4} \xi_1^2 \right) \exp \left[-\frac{5}{8} \xi_1^2  \right]+ \left(-\frac{8}{5} + \frac{1}{2} \xi_1^2  \right) \exp\left[ - \frac{5}{2} \xi_1^2  \right]\right\} .
\end{align}
Taking a high-peak approximation $\gamma \nu \gg 1 $ and $\gamma \simeq 1$, one can obtain an analytic form of the $\beta_\text{PBH} (M_{\text{PBH}})$ associated with the PBH mass~\cite{Green:2004wb}
\begin{align}
\beta_\text{PBH} (M_\text{PBH})  = \frac{1}{\sqrt{2\pi}}\left( \frac{R ~  \sigma_1}{\sqrt{3} \sigma_0 }\right)^3 \left(\nu_c^2 -1 \right)\exp \left(- \frac{\nu_c^2}{2}  \right) .
\end{align}
The current day PBH energy density fraction against the total dark matter energy density $f_\text{PBH} (M_{\text{PBH}})$ is obtained from $\beta_{\text{PBH}} (M_{\text{PBH}}) $ through the following relation
\begin{align}
f_\text{PBH}\left(M_\text{PBH}\right) 
\equiv \left. \frac{\Omega_{\rm PBH}}{\Omega_{\rm DM}} \right|_\text{today}
\simeq 2.7 \times 10^8 \left(\frac{\mathcal{K}}{0.2}\right)^{\frac{1}{2}} \left(\frac{10.75}{g_*}\right)^{\frac{1}{4}}\left(\frac{M_\odot}{M_\text{PBH}}\right)^{\frac{1}{2}} \beta_\text{PBH} (M_\text{PBH}) . 
\end{align}

\bibliographystyle{JHEP}
\bibliography{JCAP}

\providecommand{\noopsort}[1]{}\providecommand{\singleletter}[1]{#1}%

\providecommand{\href}[2]{#2}\begingroup\raggedright\begin{thebibliography}{100}

\bibitem{LIGOScientific:2016aoc}
{\scshape LIGO Scientific, Virgo} collaboration, \emph{{Observation of
  Gravitational Waves from a Binary Black Hole Merger}},
  \href{https://doi.org/10.1103/PhysRevLett.116.061102}{\emph{Phys. Rev. Lett.}
  {\bfseries 116} (2016) 061102}
  [\href{https://arxiv.org/abs/1602.03837}{{\ttfamily 1602.03837}}].

\bibitem{LIGOScientific:2021djp}
{\scshape LIGO Scientific, VIRGO, KAGRA} collaboration, \emph{{GWTC-3: Compact
  Binary Coalescences Observed by LIGO and Virgo During the Second Part of the
  Third Observing Run}},  \href{https://arxiv.org/abs/2111.03606}{{\ttfamily
  2111.03606}}.

\bibitem{NANOGrav:2020bcs}
{\scshape NANOGrav} collaboration, \emph{{The NANOGrav 12.5 yr Data Set: Search
  for an Isotropic Stochastic Gravitational-wave Background}},
  \href{https://doi.org/10.3847/2041-8213/abd401}{\emph{Astrophys. J. Lett.}
  {\bfseries 905} (2020) L34}
  [\href{https://arxiv.org/abs/2009.04496}{{\ttfamily 2009.04496}}].

\bibitem{Goncharov:2021oub}
B.~Goncharov et~al., \emph{{On the Evidence for a Common-spectrum Process in
  the Search for the Nanohertz Gravitational-wave Background with the Parkes
  Pulsar Timing Array}},
  \href{https://doi.org/10.3847/2041-8213/ac17f4}{\emph{Astrophys. J. Lett.}
  {\bfseries 917} (2021) L19}
  [\href{https://arxiv.org/abs/2107.12112}{{\ttfamily 2107.12112}}].

\bibitem{Wu:2021kmd}
Y.-M.~Wu, Z.-C.~Chen and Q.-G.~Huang, \emph{{Constraining the Polarization of
  Gravitational Waves with the Parkes Pulsar Timing Array Second Data
  Release}}, \href{https://doi.org/10.3847/1538-4357/ac35cc}{\emph{Astrophys.
  J.} {\bfseries 925} (2022) 37}
  [\href{https://arxiv.org/abs/2108.10518}{{\ttfamily 2108.10518}}].

\bibitem{NANOGrav:2021ini}
{\scshape NANOGrav} collaboration, \emph{{The NANOGrav 12.5-year Data Set:
  Search for Non-Einsteinian Polarization Modes in the Gravitational-wave
  Background}},
  \href{https://doi.org/10.3847/2041-8213/ac401c}{\emph{Astrophys. J. Lett.}
  {\bfseries 923} (2021) L22}
  [\href{https://arxiv.org/abs/2109.14706}{{\ttfamily 2109.14706}}].

\bibitem{Xue:2021gyq}
X.~Xue et~al., \emph{{Constraining Cosmological Phase Transitions with the
  Parkes Pulsar Timing Array}},
  \href{https://doi.org/10.1103/PhysRevLett.127.251303}{\emph{Phys. Rev. Lett.}
  {\bfseries 127} (2021) 251303}
  [\href{https://arxiv.org/abs/2110.03096}{{\ttfamily 2110.03096}}].

\bibitem{Antoniadis:2022pcn}
J.~Antoniadis et~al., \emph{{The International Pulsar Timing Array second data
  release: Search for an isotropic gravitational wave background}},
  \href{https://doi.org/10.1093/mnras/stab3418}{\emph{Mon. Not. Roy. Astron.
  Soc.} {\bfseries 510} (2022) 4873}
  [\href{https://arxiv.org/abs/2201.03980}{{\ttfamily 2201.03980}}].

\bibitem{Vaskonen:2020lbd}
V.~Vaskonen and H.~Veerm\"ae, \emph{{Did NANOGrav see a signal from primordial
  black hole formation?}},
  \href{https://doi.org/10.1103/PhysRevLett.126.051303}{\emph{Phys. Rev. Lett.}
  {\bfseries 126} (2021) 051303}
  [\href{https://arxiv.org/abs/2009.07832}{{\ttfamily 2009.07832}}].

\bibitem{DeLuca:2020agl}
V.~De~Luca, G.~Franciolini and A.~Riotto, \emph{{NANOGrav Data Hints at
  Primordial Black Holes as Dark Matter}},
  \href{https://doi.org/10.1103/PhysRevLett.126.041303}{\emph{Phys. Rev. Lett.}
  {\bfseries 126} (2021) 041303}
  [\href{https://arxiv.org/abs/2009.08268}{{\ttfamily 2009.08268}}].

\bibitem{Kohri:2020qqd}
K.~Kohri and T.~Terada, \emph{{Solar-Mass Primordial Black Holes Explain
  NANOGrav Hint of Gravitational Waves}},
  \href{https://doi.org/10.1016/j.physletb.2020.136040}{\emph{Phys. Lett. B}
  {\bfseries 813} (2021) 136040}
  [\href{https://arxiv.org/abs/2009.11853}{{\ttfamily 2009.11853}}].

\bibitem{Domenech:2020ers}
G.~Dom\`enech and S.~Pi, \emph{{NANOGrav hints on planet-mass primordial black
  holes}}, \href{https://doi.org/10.1007/s11433-021-1839-6}{\emph{Sci. China
  Phys. Mech. Astron.} {\bfseries 65} (2022) 230411}
  [\href{https://arxiv.org/abs/2010.03976}{{\ttfamily 2010.03976}}].

\bibitem{Baker:2019nia}
J.~Baker et~al., \emph{{The Laser Interferometer Space Antenna: Unveiling the
  Millihertz Gravitational Wave Sky}},
  \href{https://arxiv.org/abs/1907.06482}{{\ttfamily 1907.06482}}.

\bibitem{LISACosmologyWorkingGroup:2022jok}
{\scshape LISA Cosmology Working Group} collaboration, \emph{{Cosmology with
  the Laser Interferometer Space Antenna}},
  \href{https://arxiv.org/abs/2204.05434}{{\ttfamily 2204.05434}}.

\bibitem{Sato:2017dkf}
S.~Sato et~al., \emph{{The status of DECIGO}},
  \href{https://doi.org/10.1088/1742-6596/840/1/012010}{\emph{J. Phys. Conf.
  Ser.} {\bfseries 840} (2017) 012010}.

\bibitem{Kawamura:2020pcg}
S.~Kawamura et~al., \emph{{Current status of space gravitational wave antenna
  DECIGO and B-DECIGO}},
  \href{https://doi.org/10.1093/ptep/ptab019}{\emph{PTEP} {\bfseries 2021}
  (2021) 05A105} [\href{https://arxiv.org/abs/2006.13545}{{\ttfamily
  2006.13545}}].

\bibitem{Maggiore:2019uih}
M.~Maggiore et~al., \emph{{Science Case for the Einstein Telescope}},
  \href{https://doi.org/10.1088/1475-7516/2020/03/050}{\emph{JCAP} {\bfseries
  03} (2020) 050} [\href{https://arxiv.org/abs/1912.02622}{{\ttfamily
  1912.02622}}].

\bibitem{Weltman:2018zrl}
A.~Weltman et~al., \emph{{Fundamental physics with the Square Kilometre
  Array}}, \href{https://doi.org/10.1017/pasa.2019.42}{\emph{Publ. Astron. Soc.
  Austral.} {\bfseries 37} (2020) e002}
  [\href{https://arxiv.org/abs/1810.02680}{{\ttfamily 1810.02680}}].

\bibitem{Sedda:2019uro}
M.A.~Sedda et~al., \emph{{The missing link in gravitational-wave astronomy:
  discoveries waiting in the decihertz range}},
  \href{https://doi.org/10.1088/1361-6382/abb5c1}{\emph{Class. Quant. Grav.}
  {\bfseries 37} (2020) 215011}
  [\href{https://arxiv.org/abs/1908.11375}{{\ttfamily 1908.11375}}].

\bibitem{Caldwell:2022qsj}
R.~Caldwell et~al., \emph{{Detection of Early-Universe Gravitational Wave
  Signatures and Fundamental Physics}},
  \href{https://arxiv.org/abs/2203.07972}{{\ttfamily 2203.07972}}.

\bibitem{Domenech:2021ztg}
G.~Dom\`enech, \emph{{Scalar Induced Gravitational Waves Review}},
  \href{https://doi.org/10.3390/universe7110398}{\emph{Universe} {\bfseries 7}
  (2021) 398} [\href{https://arxiv.org/abs/2109.01398}{{\ttfamily
  2109.01398}}].

\bibitem{Ivanov:1994pa}
P.~Ivanov, P.~Naselsky and I.~Novikov, \emph{{Inflation and primordial black
  holes as dark matter}},
  \href{https://doi.org/10.1103/PhysRevD.50.7173}{\emph{Phys. Rev.} {\bfseries
  D50} (1994) 7173}.

\bibitem{Green:1997sz}
A.M.~Green and A.R.~Liddle, \emph{{Constraints on the density perturbation
  spectrum from primordial black holes}},
  \href{https://doi.org/10.1103/PhysRevD.56.6166}{\emph{Phys. Rev.} {\bfseries
  D56} (1997) 6166} [\href{https://arxiv.org/abs/astro-ph/9704251}{{\ttfamily
  astro-ph/9704251}}].

\bibitem{Drees:2011hb}
M.~Drees and E.~Erfani, \emph{{Running-Mass Inflation Model and Primordial
  Black Holes}},
  \href{https://doi.org/10.1088/1475-7516/2011/04/005}{\emph{JCAP} {\bfseries
  1104} (2011) 005} [\href{https://arxiv.org/abs/1102.2340}{{\ttfamily
  1102.2340}}].

\bibitem{Kohri:2012yw}
K.~Kohri, C.-M.~Lin and T.~Matsuda, \emph{{Primordial black holes from the
  inflating curvaton}},
  \href{https://doi.org/10.1103/PhysRevD.87.103527}{\emph{Phys. Rev.}
  {\bfseries D87} (2013) 103527}
  [\href{https://arxiv.org/abs/1211.2371}{{\ttfamily 1211.2371}}].

\bibitem{Clesse:2015wea}
S.~Clesse and J.~Garc{\'\i}a-Bellido, \emph{{Massive Primordial Black Holes
  from Hybrid Inflation as Dark Matter and the seeds of Galaxies}},
  \href{https://doi.org/10.1103/PhysRevD.92.023524}{\emph{Phys. Rev.}
  {\bfseries D92} (2015) 023524}
  [\href{https://arxiv.org/abs/1501.07565}{{\ttfamily 1501.07565}}].

\bibitem{Nakama:2016gzw}
T.~Nakama, J.~Silk and M.~Kamionkowski, \emph{{Stochastic gravitational waves
  associated with the formation of primordial black holes}},
  \href{https://doi.org/10.1103/PhysRevD.95.043511}{\emph{Phys. Rev. D}
  {\bfseries 95} (2017) 043511}
  [\href{https://arxiv.org/abs/1612.06264}{{\ttfamily 1612.06264}}].

\bibitem{Inomata:2017okj}
K.~Inomata, M.~Kawasaki, K.~Mukaida, Y.~Tada and T.T.~Yanagida,
  \emph{{Inflationary Primordial Black Holes as All Dark Matter}},
  \href{https://doi.org/10.1103/PhysRevD.96.043504}{\emph{Phys. Rev. D}
  {\bfseries 96} (2017) 043504}
  [\href{https://arxiv.org/abs/1701.02544}{{\ttfamily 1701.02544}}].

\bibitem{Garcia-Bellido:2017mdw}
J.~Garcia-Bellido and E.~Ruiz~Morales, \emph{{Primordial black holes from
  single field models of inflation}},
  \href{https://doi.org/10.1016/j.dark.2017.09.007}{\emph{Phys. Dark Univ.}
  {\bfseries 18} (2017) 47} [\href{https://arxiv.org/abs/1702.03901}{{\ttfamily
  1702.03901}}].

\bibitem{Domcke:2017fix}
V.~Domcke, F.~Muia, M.~Pieroni and L.T.~Witkowski, \emph{{PBH dark matter from
  axion inflation}},
  \href{https://doi.org/10.1088/1475-7516/2017/07/048}{\emph{JCAP} {\bfseries
  07} (2017) 048} [\href{https://arxiv.org/abs/1704.03464}{{\ttfamily
  1704.03464}}].

\bibitem{Kannike:2017bxn}
K.~Kannike, L.~Marzola, M.~Raidal and H.~Veerm{\"a}e, \emph{{Single Field
  Double Inflation and Primordial Black Holes}},
  \href{https://doi.org/10.1088/1475-7516/2017/09/020}{\emph{JCAP} {\bfseries
  1709} (2017) 020} [\href{https://arxiv.org/abs/1705.06225}{{\ttfamily
  1705.06225}}].

\bibitem{Carr:2017edp}
B.~Carr, T.~Tenkanen and V.~Vaskonen, \emph{{Primordial black holes from
  inflaton and spectator field perturbations in a matter-dominated era}},
  \href{https://doi.org/10.1103/PhysRevD.96.063507}{\emph{Phys. Rev. D}
  {\bfseries 96} (2017) 063507}
  [\href{https://arxiv.org/abs/1706.03746}{{\ttfamily 1706.03746}}].

\bibitem{Germani:2017bcs}
C.~Germani and T.~Prokopec, \emph{{On primordial black holes from an inflection
  point}}, \href{https://doi.org/10.1016/j.dark.2017.09.001}{\emph{Phys. Dark
  Univ.} {\bfseries 18} (2017) 6}
  [\href{https://arxiv.org/abs/1706.04226}{{\ttfamily 1706.04226}}].

\bibitem{Motohashi:2017kbs}
H.~Motohashi and W.~Hu, \emph{{Primordial Black Holes and Slow-Roll
  Violation}}, \href{https://doi.org/10.1103/PhysRevD.96.063503}{\emph{Phys.
  Rev.} {\bfseries D96} (2017) 063503}
  [\href{https://arxiv.org/abs/1706.06784}{{\ttfamily 1706.06784}}].

\bibitem{Pattison:2017mbe}
C.~Pattison, V.~Vennin, H.~Assadullahi and D.~Wands, \emph{{Quantum diffusion
  during inflation and primordial black holes}},
  \href{https://doi.org/10.1088/1475-7516/2017/10/046}{\emph{JCAP} {\bfseries
  10} (2017) 046} [\href{https://arxiv.org/abs/1707.00537}{{\ttfamily
  1707.00537}}].

\bibitem{Di:2017ndc}
H.~Di and Y.~Gong, \emph{{Primordial black holes and second order gravitational
  waves from ultra-slow-roll inflation}},
  \href{https://doi.org/10.1088/1475-7516/2018/07/007}{\emph{JCAP} {\bfseries
  07} (2018) 007} [\href{https://arxiv.org/abs/1707.09578}{{\ttfamily
  1707.09578}}].

\bibitem{Inomata:2017vxo}
K.~Inomata, M.~Kawasaki, K.~Mukaida and T.T.~Yanagida, \emph{{Double inflation
  as a single origin of primordial black holes for all dark matter and LIGO
  observations}}, \href{https://doi.org/10.1103/PhysRevD.97.043514}{\emph{Phys.
  Rev.} {\bfseries D97} (2018) 043514}
  [\href{https://arxiv.org/abs/1711.06129}{{\ttfamily 1711.06129}}].

\bibitem{Ando:2017veq}
K.~Ando, K.~Inomata, M.~Kawasaki, K.~Mukaida and T.T.~Yanagida,
  \emph{{Primordial black holes for the LIGO events in the axionlike curvaton
  model}}, \href{https://doi.org/10.1103/PhysRevD.97.123512}{\emph{Phys. Rev.
  D} {\bfseries 97} (2018) 123512}
  [\href{https://arxiv.org/abs/1711.08956}{{\ttfamily 1711.08956}}].

\bibitem{Hertzberg:2017dkh}
M.P.~Hertzberg and M.~Yamada, \emph{{Primordial Black Holes from Polynomial
  Potentials in Single Field Inflation}},
  \href{https://doi.org/10.1103/PhysRevD.97.083509}{\emph{Phys. Rev. D}
  {\bfseries 97} (2018) 083509}
  [\href{https://arxiv.org/abs/1712.09750}{{\ttfamily 1712.09750}}].

\bibitem{Franciolini:2018vbk}
G.~Franciolini, A.~Kehagias, S.~Matarrese and A.~Riotto, \emph{{Primordial
  Black Holes from Inflation and non-Gaussianity}},
  \href{https://doi.org/10.1088/1475-7516/2018/03/016}{\emph{JCAP} {\bfseries
  03} (2018) 016} [\href{https://arxiv.org/abs/1801.09415}{{\ttfamily
  1801.09415}}].

\bibitem{Biagetti:2018pjj}
M.~Biagetti, G.~Franciolini, A.~Kehagias and A.~Riotto, \emph{{Primordial Black
  Holes from Inflation and Quantum Diffusion}},
  \href{https://doi.org/10.1088/1475-7516/2018/07/032}{\emph{JCAP} {\bfseries
  1807} (2018) 032} [\href{https://arxiv.org/abs/1804.07124}{{\ttfamily
  1804.07124}}].

\bibitem{Cai:2018tuh}
Y.-F.~Cai, X.~Tong, D.-G.~Wang and S.-F.~Yan, \emph{{Primordial Black Holes
  from Sound Speed Resonance during Inflation}},
  \href{https://doi.org/10.1103/PhysRevLett.121.081306}{\emph{Phys. Rev. Lett.}
  {\bfseries 121} (2018) 081306}
  [\href{https://arxiv.org/abs/1805.03639}{{\ttfamily 1805.03639}}].

\bibitem{Germani:2018jgr}
C.~Germani and I.~Musco, \emph{{Abundance of Primordial Black Holes Depends on
  the Shape of the Inflationary Power Spectrum}},
  \href{https://doi.org/10.1103/PhysRevLett.122.141302}{\emph{Phys. Rev. Lett.}
  {\bfseries 122} (2019) 141302}
  [\href{https://arxiv.org/abs/1805.04087}{{\ttfamily 1805.04087}}].

\bibitem{Dalianis:2018frf}
I.~Dalianis, A.~Kehagias and G.~Tringas, \emph{{Primordial black holes from
  \ensuremath{\alpha}-attractors}},
  \href{https://doi.org/10.1088/1475-7516/2019/01/037}{\emph{JCAP} {\bfseries
  01} (2019) 037} [\href{https://arxiv.org/abs/1805.09483}{{\ttfamily
  1805.09483}}].

\bibitem{Byrnes:2018txb}
C.T.~Byrnes, P.S.~Cole and S.P.~Patil, \emph{{Steepest growth of the power
  spectrum and primordial black holes}},
  \href{https://doi.org/10.1088/1475-7516/2019/06/028}{\emph{JCAP} {\bfseries
  06} (2019) 028} [\href{https://arxiv.org/abs/1811.11158}{{\ttfamily
  1811.11158}}].

\bibitem{Passaglia:2018ixg}
S.~Passaglia, W.~Hu and H.~Motohashi, \emph{{Primordial black holes and local
  non-Gaussianity in canonical inflation}},
  \href{https://doi.org/10.1103/PhysRevD.99.043536}{\emph{Phys. Rev.}
  {\bfseries D99} (2019) 043536}
  [\href{https://arxiv.org/abs/1812.08243}{{\ttfamily 1812.08243}}].

\bibitem{Dimopoulos:2019wew}
K.~Dimopoulos, T.~Markkanen, A.~Racioppi and V.~Vaskonen, \emph{{Primordial
  Black Holes from Thermal Inflation}},
  \href{https://doi.org/10.1088/1475-7516/2019/07/046}{\emph{JCAP} {\bfseries
  1907} (2019) 046} [\href{https://arxiv.org/abs/1903.09598}{{\ttfamily
  1903.09598}}].

\bibitem{Bhaumik:2019tvl}
N.~Bhaumik and R.K.~Jain, \emph{{Primordial black holes dark matter from
  inflection point models of inflation and the effects of reheating}},
  \href{https://doi.org/10.1088/1475-7516/2020/01/037}{\emph{JCAP} {\bfseries
  01} (2020) 037} [\href{https://arxiv.org/abs/1907.04125}{{\ttfamily
  1907.04125}}].

\bibitem{Carrilho:2019oqg}
P.~Carrilho, K.A.~Malik and D.J.~Mulryne, \emph{{Dissecting the growth of the
  power spectrum for primordial black holes}},
  \href{https://doi.org/10.1103/PhysRevD.100.103529}{\emph{Phys. Rev. D}
  {\bfseries 100} (2019) 103529}
  [\href{https://arxiv.org/abs/1907.05237}{{\ttfamily 1907.05237}}].

\bibitem{Fu:2019ttf}
C.~Fu, P.~Wu and H.~Yu, \emph{{Primordial Black Holes from Inflation with
  Nonminimal Derivative Coupling}},
  \href{https://doi.org/10.1103/PhysRevD.100.063532}{\emph{Phys. Rev. D}
  {\bfseries 100} (2019) 063532}
  [\href{https://arxiv.org/abs/1907.05042}{{\ttfamily 1907.05042}}].

\bibitem{Mishra:2019pzq}
S.S.~Mishra and V.~Sahni, \emph{{Primordial Black Holes from a tiny bump/dip in
  the Inflaton potential}},
  \href{https://doi.org/10.1088/1475-7516/2020/04/007}{\emph{JCAP} {\bfseries
  04} (2020) 007} [\href{https://arxiv.org/abs/1911.00057}{{\ttfamily
  1911.00057}}].

\bibitem{Cai:2019bmk}
R.-G.~Cai, Z.-K.~Guo, J.~Liu, L.~Liu and X.-Y.~Yang, \emph{{Primordial black
  holes and gravitational waves from parametric amplification of curvature
  perturbations}},
  \href{https://doi.org/10.1088/1475-7516/2020/06/013}{\emph{JCAP} {\bfseries
  06} (2020) 013} [\href{https://arxiv.org/abs/1912.10437}{{\ttfamily
  1912.10437}}].

\bibitem{Cheong:2019vzl}
D.Y.~Cheong, S.M.~Lee and S.C.~Park, \emph{{Primordial black holes in
  Higgs-$R^2$ inflation as the whole of dark matter}},
  \href{https://doi.org/10.1088/1475-7516/2021/01/032}{\emph{JCAP} {\bfseries
  01} (2021) 032} [\href{https://arxiv.org/abs/1912.12032}{{\ttfamily
  1912.12032}}].

\bibitem{Ashoorioon:2019xqc}
A.~Ashoorioon, A.~Rostami and J.T.~Firouzjaee, \emph{{EFT compatible PBHs:
  effective spawning of the seeds for primordial black holes during
  inflation}}, \href{https://doi.org/10.1007/JHEP07(2021)087}{\emph{JHEP}
  {\bfseries 07} (2021) 087}
  [\href{https://arxiv.org/abs/1912.13326}{{\ttfamily 1912.13326}}].

\bibitem{Lin:2020goi}
J.~Lin, Q.~Gao, Y.~Gong, Y.~Lu, C.~Zhang and F.~Zhang, \emph{{Primordial black
  holes and secondary gravitational waves from $k$ and $G$ inflation}},
  \href{https://doi.org/10.1103/PhysRevD.101.103515}{\emph{Phys. Rev. D}
  {\bfseries 101} (2020) 103515}
  [\href{https://arxiv.org/abs/2001.05909}{{\ttfamily 2001.05909}}].

\bibitem{Ballesteros:2020qam}
G.~Ballesteros, J.~Rey, M.~Taoso and A.~Urbano, \emph{{Primordial black holes
  as dark matter and gravitational waves from single-field polynomial
  inflation}}, \href{https://doi.org/10.1088/1475-7516/2020/07/025}{\emph{JCAP}
  {\bfseries 07} (2020) 025}
  [\href{https://arxiv.org/abs/2001.08220}{{\ttfamily 2001.08220}}].

\bibitem{Palma:2020ejf}
G.A.~Palma, S.~Sypsas and C.~Zenteno, \emph{{Seeding primordial black holes in
  multifield inflation}},
  \href{https://doi.org/10.1103/PhysRevLett.125.121301}{\emph{Phys. Rev. Lett.}
  {\bfseries 125} (2020) 121301}
  [\href{https://arxiv.org/abs/2004.06106}{{\ttfamily 2004.06106}}].

\bibitem{Fumagalli:2020adf}
J.~Fumagalli, S.~Renaux-Petel, J.W.~Ronayne and L.T.~Witkowski, \emph{{Turning
  in the landscape: a new mechanism for generating Primordial Black Holes}},
  \href{https://arxiv.org/abs/2004.08369}{{\ttfamily 2004.08369}}.

\bibitem{Braglia:2020eai}
M.~Braglia, D.K.~Hazra, F.~Finelli, G.F.~Smoot, L.~Sriramkumar and
  A.A.~Starobinsky, \emph{{Generating PBHs and small-scale GWs in two-field
  models of inflation}},
  \href{https://doi.org/10.1088/1475-7516/2020/08/001}{\emph{JCAP} {\bfseries
  08} (2020) 001} [\href{https://arxiv.org/abs/2005.02895}{{\ttfamily
  2005.02895}}].

\bibitem{Aldabergenov:2020bpt}
Y.~Aldabergenov, A.~Addazi and S.V.~Ketov, \emph{{Primordial black holes from
  modified supergravity}},
  \href{https://doi.org/10.1140/epjc/s10052-020-08506-6}{\emph{Eur. Phys. J. C}
  {\bfseries 80} (2020) 917}
  [\href{https://arxiv.org/abs/2006.16641}{{\ttfamily 2006.16641}}].

\bibitem{Aldabergenov:2020yok}
Y.~Aldabergenov, A.~Addazi and S.V.~Ketov, \emph{{Testing Primordial Black
  Holes as Dark Matter in Supergravity from Gravitational Waves}},
  \href{https://doi.org/10.1016/j.physletb.2021.136069}{\emph{Phys. Lett. B}
  {\bfseries 814} (2021) 136069}
  [\href{https://arxiv.org/abs/2008.10476}{{\ttfamily 2008.10476}}].

\bibitem{Gundhi:2020kzm}
A.~Gundhi, S.V.~Ketov and C.F.~Steinwachs, \emph{{Primordial black hole dark
  matter in dilaton-extended two-field Starobinsky inflation}},
  \href{https://doi.org/10.1103/PhysRevD.103.083518}{\emph{Phys. Rev. D}
  {\bfseries 103} (2021) 083518}
  [\href{https://arxiv.org/abs/2011.05999}{{\ttfamily 2011.05999}}].

\bibitem{Gundhi:2020zvb}
A.~Gundhi and C.F.~Steinwachs, \emph{{Scalaron\textendash{}Higgs inflation
  reloaded: Higgs-dependent scalaron mass and primordial black hole dark
  matter}}, \href{https://doi.org/10.1140/epjc/s10052-021-09225-2}{\emph{Eur.
  Phys. J. C} {\bfseries 81} (2021) 460}
  [\href{https://arxiv.org/abs/2011.09485}{{\ttfamily 2011.09485}}].

\bibitem{Zheng:2021vda}
R.~Zheng, J.~Shi and T.~Qiu, \emph{{On Primordial Black Holes generated from
  inflation with solo/multi-bumpy potential}},
  \href{https://arxiv.org/abs/2106.04303}{{\ttfamily 2106.04303}}.

\bibitem{Chen:2021nio}
P.~Chen, S.~Koh and G.~Tumurtushaa, \emph{{Primordial black holes and induced
  gravitational waves from inflation in the Horndeski theory of gravity}},
  \href{https://arxiv.org/abs/2107.08638}{{\ttfamily 2107.08638}}.

\bibitem{Kawai:2021edk}
S.~Kawai and J.~Kim, \emph{{Primordial black holes from Gauss-Bonnet-corrected
  single field inflation}},
  \href{https://doi.org/10.1103/PhysRevD.104.083545}{\emph{Phys. Rev. D}
  {\bfseries 104} (2021) 083545}
  [\href{https://arxiv.org/abs/2108.01340}{{\ttfamily 2108.01340}}].

\bibitem{Rezazadeh:2021clf}
K.~Rezazadeh, Z.~Teimoori and K.~Karami, \emph{{Non-Gaussianity and Secondary
  Gravitational Waves from Primordial Black Holes Production in
  $\alpha$-attractor Inflation}},
  \href{https://arxiv.org/abs/2110.01482}{{\ttfamily 2110.01482}}.

\bibitem{Iacconi:2021ltm}
L.~Iacconi, H.~Assadullahi, M.~Fasiello and D.~Wands, \emph{{Revisiting
  small-scale fluctuations in $\alpha$-attractor models of inflation}},
  \href{https://arxiv.org/abs/2112.05092}{{\ttfamily 2112.05092}}.

\bibitem{Pi:2021dft}
S.~Pi and M.~Sasaki, \emph{{Primordial Black Hole Formation in Non-Minimal
  Curvaton Scenario}},  \href{https://arxiv.org/abs/2112.12680}{{\ttfamily
  2112.12680}}.

\bibitem{Papanikolaou:2021uhe}
T.~Papanikolaou, C.~Tzerefos, S.~Basilakos and E.N.~Saridakis, \emph{{Scalar
  induced gravitational waves from primordial black hole Poisson fluctuations
  in Starobinsky inflation}},
  \href{https://arxiv.org/abs/2112.15059}{{\ttfamily 2112.15059}}.

\bibitem{Ashoorioon:2022raz}
A.~Ashoorioon, K.~Rezazadeh and A.~Rostami, \emph{{NANOGrav Signal from the End
  of Inflation and the LIGO Mass and Heavier Primordial Black Holes}},
  \href{https://arxiv.org/abs/2202.01131}{{\ttfamily 2202.01131}}.

\bibitem{Kallosh:2022vha}
R.~Kallosh and A.~Linde, \emph{{Dilaton-Axion Inflation with PBHs and GWs}},
  \href{https://arxiv.org/abs/2203.10437}{{\ttfamily 2203.10437}}.

\bibitem{Geller:2022nkr}
S.~Geller, W.~Qin, E.~McDonough and D.I.~Kaiser, \emph{{Primordial Black Holes
  from Multifield Inflation with Nonminimal Couplings}},
  \href{https://arxiv.org/abs/2205.04471}{{\ttfamily 2205.04471}}.

\bibitem{Karam:2022nym}
A.~Karam, N.~Koivunen, E.~Tomberg, V.~Vaskonen and H.~Veerm\"ae, \emph{{Anatomy
  of single-field inflationary models for primordial black holes}},
  \href{https://arxiv.org/abs/2205.13540}{{\ttfamily 2205.13540}}.

\bibitem{Cai:2018dig}
R.-G.~Cai, S.~Pi and M.~Sasaki, \emph{{Gravitational Waves Induced by
  non-Gaussian Scalar Perturbations}},
  \href{https://doi.org/10.1103/PhysRevLett.122.201101}{\emph{Phys. Rev. Lett.}
  {\bfseries 122} (2019) 201101}
  [\href{https://arxiv.org/abs/1810.11000}{{\ttfamily 1810.11000}}].

\bibitem{Bartolo:2018evs}
N.~Bartolo, V.~De~Luca, G.~Franciolini, A.~Lewis, M.~Peloso and A.~Riotto,
  \emph{{Primordial Black Hole Dark Matter: LISA Serendipity}},
  \href{https://doi.org/10.1103/PhysRevLett.122.211301}{\emph{Phys. Rev. Lett.}
  {\bfseries 122} (2019) 211301}
  [\href{https://arxiv.org/abs/1810.12218}{{\ttfamily 1810.12218}}].

\bibitem{Braglia:2020taf}
M.~Braglia, X.~Chen and D.K.~Hazra, \emph{{Probing Primordial Features with the
  Stochastic Gravitational Wave Background}},
  \href{https://doi.org/10.1088/1475-7516/2021/03/005}{\emph{JCAP} {\bfseries
  03} (2021) 005} [\href{https://arxiv.org/abs/2012.05821}{{\ttfamily
  2012.05821}}].

\bibitem{Braglia:2021wwa}
M.~Braglia, J.~Garcia-Bellido and S.~Kuroyanagi, \emph{{Testing Primordial
  Black Holes with multi-band observations of the stochastic gravitational wave
  background}},
  \href{https://doi.org/10.1088/1475-7516/2021/12/012}{\emph{JCAP} {\bfseries
  12} (2021) 012} [\href{https://arxiv.org/abs/2110.07488}{{\ttfamily
  2110.07488}}].

\bibitem{Carr:2020gox}
B.~Carr, K.~Kohri, Y.~Sendouda and J.~Yokoyama, \emph{{Constraints on
  primordial black holes}},
  \href{https://doi.org/10.1088/1361-6633/ac1e31}{\emph{Rept. Prog. Phys.}
  {\bfseries 84} (2021) 116902}
  [\href{https://arxiv.org/abs/2002.12778}{{\ttfamily 2002.12778}}].

\bibitem{Bezrukov:2007ep}
F.L.~Bezrukov and M.~Shaposhnikov, \emph{{The Standard Model Higgs boson as the
  inflaton}}, \href{https://doi.org/10.1016/j.physletb.2007.11.072}{\emph{Phys.
  Lett.} {\bfseries B659} (2008) 703}
  [\href{https://arxiv.org/abs/0710.3755}{{\ttfamily 0710.3755}}].

\bibitem{Giudice:2010ka}
G.F.~Giudice and H.M.~Lee, \emph{{Unitarizing Higgs Inflation}},
  \href{https://doi.org/10.1016/j.physletb.2010.10.035}{\emph{Phys. Lett.}
  {\bfseries B694} (2011) 294}
  [\href{https://arxiv.org/abs/1010.1417}{{\ttfamily 1010.1417}}].

\bibitem{Bezrukov:2010jz}
F.~Bezrukov, A.~Magnin, M.~Shaposhnikov and S.~Sibiryakov, \emph{{Higgs
  inflation: consistency and generalisations}},
  \href{https://doi.org/10.1007/JHEP01(2011)016}{\emph{JHEP} {\bfseries 01}
  (2011) 016} [\href{https://arxiv.org/abs/1008.5157}{{\ttfamily 1008.5157}}].

\bibitem{Lerner:2011it}
R.N.~Lerner and J.~McDonald, \emph{{Unitarity-Violation in Generalized Higgs
  Inflation Models}},
  \href{https://doi.org/10.1088/1475-7516/2012/11/019}{\emph{JCAP} {\bfseries
  1211} (2012) 019} [\href{https://arxiv.org/abs/1112.0954}{{\ttfamily
  1112.0954}}].

\bibitem{Calmet:2013hia}
X.~Calmet and R.~Casadio, \emph{{Self-healing of unitarity in Higgs
  inflation}},
  \href{https://doi.org/10.1016/j.physletb.2014.05.008}{\emph{Phys. Lett.}
  {\bfseries B734} (2014) 17}
  [\href{https://arxiv.org/abs/1310.7410}{{\ttfamily 1310.7410}}].

\bibitem{Hamada:2014iga}
Y.~Hamada, H.~Kawai, K.-y.~Oda and S.C.~Park, \emph{{Higgs Inflation is Still
  Alive after the Results from BICEP2}},
  \href{https://doi.org/10.1103/PhysRevLett.112.241301}{\emph{Phys. Rev. Lett.}
  {\bfseries 112} (2014) 241301}
  [\href{https://arxiv.org/abs/1403.5043}{{\ttfamily 1403.5043}}].

\bibitem{Bezrukov:2014bra}
F.~Bezrukov and M.~Shaposhnikov, \emph{{Higgs inflation at the critical
  point}}, \href{https://doi.org/10.1016/j.physletb.2014.05.074}{\emph{Phys.
  Lett.} {\bfseries B734} (2014) 249}
  [\href{https://arxiv.org/abs/1403.6078}{{\ttfamily 1403.6078}}].

\bibitem{Herranen:2014cua}
M.~Herranen, T.~Markkanen, S.~Nurmi and A.~Rajantie, \emph{{Spacetime curvature
  and the Higgs stability during inflation}},
  \href{https://doi.org/10.1103/PhysRevLett.113.211102}{\emph{Phys. Rev. Lett.}
  {\bfseries 113} (2014) 211102}
  [\href{https://arxiv.org/abs/1407.3141}{{\ttfamily 1407.3141}}].

\bibitem{Hamada:2014wna}
Y.~Hamada, H.~Kawai, K.-y.~Oda and S.C.~Park, \emph{{Higgs inflation from
  Standard Model criticality}},
  \href{https://doi.org/10.1103/PhysRevD.91.053008}{\emph{Phys. Rev.}
  {\bfseries D91} (2015) 053008}
  [\href{https://arxiv.org/abs/1408.4864}{{\ttfamily 1408.4864}}].

\bibitem{Bezrukov:2014ipa}
F.~Bezrukov, J.~Rubio and M.~Shaposhnikov, \emph{{Living beyond the edge: Higgs
  inflation and vacuum metastability}},
  \href{https://doi.org/10.1103/PhysRevD.92.083512}{\emph{Phys. Rev. D}
  {\bfseries 92} (2015) 083512}
  [\href{https://arxiv.org/abs/1412.3811}{{\ttfamily 1412.3811}}].

\bibitem{Barbon:2015fla}
J.L.F.~Barbon, J.A.~Casas, J.~Elias-Miro and J.R.~Espinosa, \emph{{Higgs
  Inflation as a Mirage}},
  \href{https://doi.org/10.1007/JHEP09(2015)027}{\emph{JHEP} {\bfseries 09}
  (2015) 027} [\href{https://arxiv.org/abs/1501.02231}{{\ttfamily
  1501.02231}}].

\bibitem{Salvio:2015kka}
A.~Salvio and A.~Mazumdar, \emph{{Classical and Quantum Initial Conditions for
  Higgs Inflation}},
  \href{https://doi.org/10.1016/j.physletb.2015.09.020}{\emph{Phys. Lett.}
  {\bfseries B750} (2015) 194}
  [\href{https://arxiv.org/abs/1506.07520}{{\ttfamily 1506.07520}}].

\bibitem{Calmet:2016fsr}
X.~Calmet and I.~Kuntz, \emph{{Higgs Starobinsky Inflation}},
  \href{https://doi.org/10.1140/epjc/s10052-016-4136-3}{\emph{Eur. Phys. J.}
  {\bfseries C76} (2016) 289}
  [\href{https://arxiv.org/abs/1605.02236}{{\ttfamily 1605.02236}}].

\bibitem{Hamada:2016onh}
Y.~Hamada, H.~Kawai, Y.~Nakanishi and K.-y.~Oda, \emph{{Meaning of the field
  dependence of the renormalization scale in Higgs inflation}},
  \href{https://doi.org/10.1103/PhysRevD.95.103524}{\emph{Phys. Rev. D}
  {\bfseries 95} (2017) 103524}
  [\href{https://arxiv.org/abs/1610.05885}{{\ttfamily 1610.05885}}].

\bibitem{Ema:2017rqn}
Y.~Ema, \emph{{Higgs Scalaron Mixed Inflation}},
  \href{https://doi.org/10.1016/j.physletb.2017.04.060}{\emph{Phys. Lett.}
  {\bfseries B770} (2017) 403}
  [\href{https://arxiv.org/abs/1701.07665}{{\ttfamily 1701.07665}}].

\bibitem{Ezquiaga:2017fvi}
J.M.~Ezquiaga, J.~Garcia-Bellido and E.~Ruiz~Morales, \emph{{Primordial Black
  Hole production in Critical Higgs Inflation}},
  \href{https://doi.org/10.1016/j.physletb.2017.11.039}{\emph{Phys. Lett. B}
  {\bfseries 776} (2018) 345}
  [\href{https://arxiv.org/abs/1705.04861}{{\ttfamily 1705.04861}}].

\bibitem{Bezrukov:2017dyv}
F.~Bezrukov, M.~Pauly and J.~Rubio, \emph{{On the robustness of the primordial
  power spectrum in renormalized Higgs inflation}},
  \href{https://doi.org/10.1088/1475-7516/2018/02/040}{\emph{JCAP} {\bfseries
  1802} (2018) 040} [\href{https://arxiv.org/abs/1706.05007}{{\ttfamily
  1706.05007}}].

\bibitem{Hamada:2017sga}
Y.~Hamada, H.~Kawai, Y.~Nakanishi and K.-y.~Oda, \emph{{Cosmological
  implications of Standard Model criticality and Higgs inflation}},
  \href{https://doi.org/10.1016/j.nuclphysb.2020.114946}{\emph{Nucl. Phys. B}
  {\bfseries 953} (2020) 114946}
  [\href{https://arxiv.org/abs/1709.09350}{{\ttfamily 1709.09350}}].

\bibitem{Lee:2018esk}
H.M.~Lee, \emph{{Light inflaton completing Higgs inflation}},
  \href{https://doi.org/10.1103/PhysRevD.98.015020}{\emph{Phys. Rev.}
  {\bfseries D98} (2018) 015020}
  [\href{https://arxiv.org/abs/1802.06174}{{\ttfamily 1802.06174}}].

\bibitem{He:2018gyf}
M.~He, A.A.~Starobinsky and J.~Yokoyama, \emph{{Inflation in the mixed
  Higgs-$R^2$ model}},
  \href{https://doi.org/10.1088/1475-7516/2018/05/064}{\emph{JCAP} {\bfseries
  1805} (2018) 064} [\href{https://arxiv.org/abs/1804.00409}{{\ttfamily
  1804.00409}}].

\bibitem{Masina:2018ejw}
I.~Masina, \emph{{Ruling out Critical Higgs Inflation?}},
  \href{https://doi.org/10.1103/PhysRevD.98.043536}{\emph{Phys. Rev.}
  {\bfseries D98} (2018) 043536}
  [\href{https://arxiv.org/abs/1805.02160}{{\ttfamily 1805.02160}}].

\bibitem{Gorbunov:2018llf}
D.~Gorbunov and A.~Tokareva, \emph{{Scalaron the healer: removing the
  strong-coupling in the Higgs- and Higgs-dilaton inflations}},
  \href{https://doi.org/10.1016/j.physletb.2018.11.015}{\emph{Phys. Lett.}
  {\bfseries B788} (2019) 37}
  [\href{https://arxiv.org/abs/1807.02392}{{\ttfamily 1807.02392}}].

\bibitem{Ghilencea:2018rqg}
D.M.~Ghilencea, \emph{{Two-loop corrections to Starobinsky-Higgs inflation}},
  \href{https://doi.org/10.1103/PhysRevD.98.103524}{\emph{Phys. Rev.}
  {\bfseries D98} (2018) 103524}
  [\href{https://arxiv.org/abs/1807.06900}{{\ttfamily 1807.06900}}].

\bibitem{Gundhi:2018wyz}
A.~Gundhi and C.F.~Steinwachs, \emph{{Scalaron-Higgs inflation}},
  \href{https://doi.org/10.1016/j.nuclphysb.2020.114989}{\emph{Nucl. Phys. B}
  {\bfseries 954} (2020) 114989}
  [\href{https://arxiv.org/abs/1810.10546}{{\ttfamily 1810.10546}}].

\bibitem{Rasanen:2018fom}
S.~Rasanen and E.~Tomberg, \emph{{Planck scale black hole dark matter from
  Higgs inflation}},
  \href{https://doi.org/10.1088/1475-7516/2019/01/038}{\emph{JCAP} {\bfseries
  01} (2019) 038} [\href{https://arxiv.org/abs/1810.12608}{{\ttfamily
  1810.12608}}].

\bibitem{He:2018mgb}
M.~He, R.~Jinno, K.~Kamada, S.C.~Park, A.A.~Starobinsky and J.~Yokoyama,
  \emph{{On the violent preheating in the mixed Higgs-$R^2$ inflationary
  model}}, \href{https://doi.org/10.1016/j.physletb.2019.02.008}{\emph{Phys.
  Lett.} {\bfseries B791} (2019) 36}
  [\href{https://arxiv.org/abs/1812.10099}{{\ttfamily 1812.10099}}].

\bibitem{Bezrukov:2019ylq}
F.~Bezrukov, D.~Gorbunov, C.~Shepherd and A.~Tokareva, \emph{{Some like it hot:
  $R^2$ heals Higgs inflation, but does not cool it}},
  \href{https://doi.org/10.1016/j.physletb.2019.06.064}{\emph{Phys. Lett. B}
  {\bfseries 795} (2019) 657}
  [\href{https://arxiv.org/abs/1904.04737}{{\ttfamily 1904.04737}}].

\bibitem{Drees:2019xpp}
M.~Drees and Y.~Xu, \emph{{Overshooting, Critical Higgs Inflation and Second
  Order Gravitational Wave Signatures}},
  \href{https://doi.org/10.1140/epjc/s10052-021-08976-2}{\emph{Eur. Phys. J. C}
  {\bfseries 81} (2021) 182}
  [\href{https://arxiv.org/abs/1905.13581}{{\ttfamily 1905.13581}}].

\bibitem{Ema:2019fdd}
Y.~Ema, \emph{{Dynamical Emergence of Scalaron in Higgs Inflation}},
  \href{https://doi.org/10.1088/1475-7516/2019/09/027}{\emph{JCAP} {\bfseries
  1909} (2019) 027} [\href{https://arxiv.org/abs/1907.00993}{{\ttfamily
  1907.00993}}].

\bibitem{Hamada:2020kuy}
Y.~Hamada, K.~Kawana and A.~Scherlis, \emph{{On Preheating in Higgs
  Inflation}}, \href{https://doi.org/10.1088/1475-7516/2021/03/062}{\emph{JCAP}
  {\bfseries 03} (2021) 062}
  [\href{https://arxiv.org/abs/2007.04701}{{\ttfamily 2007.04701}}].

\bibitem{Ema:2020evi}
Y.~Ema, K.~Mukaida and J.~van~de Vis, \emph{{Renormalization group equations of
  Higgs-R$^{2}$ inflation}},
  \href{https://doi.org/10.1007/JHEP02(2021)109}{\emph{JHEP} {\bfseries 02}
  (2021) 109} [\href{https://arxiv.org/abs/2008.01096}{{\ttfamily
  2008.01096}}].

\bibitem{Cheong:2021vdb}
D.Y.~Cheong, S.M.~Lee and S.C.~Park, \emph{{Progress in Higgs inflation}},
  \href{https://doi.org/10.1007/s40042-021-00086-2}{\emph{J. Korean Phys. Soc.}
  {\bfseries 78} (2021) 897}
  [\href{https://arxiv.org/abs/2103.00177}{{\ttfamily 2103.00177}}].

\bibitem{Lee:2021dgi}
H.M.~Lee and A.G.~Menkara, \emph{{Cosmology of linear Higgs-sigma models with
  conformal invariance}},
  \href{https://doi.org/10.1007/JHEP09(2021)018}{\emph{JHEP} {\bfseries 09}
  (2021) 018} [\href{https://arxiv.org/abs/2104.10390}{{\ttfamily
  2104.10390}}].

\bibitem{Aoki:2021aph}
S.~Aoki, H.M.~Lee and A.G.~Menkara, \emph{{Inflation and supersymmetry breaking
  in Higgs-R$^{2}$ supergravity}},
  \href{https://doi.org/10.1007/JHEP10(2021)178}{\emph{JHEP} {\bfseries 10}
  (2021) 178} [\href{https://arxiv.org/abs/2108.00222}{{\ttfamily
  2108.00222}}].

\bibitem{Lee:2021rzy}
S.M.~Lee, T.~Modak, K.-y.~Oda and T.~Takahashi, \emph{{The $R^2$-Higgs
  inflation with two Higgs doublets}},
  \href{https://doi.org/10.1140/epjc/s10052-021-09978-w}{\emph{Eur. Phys. J. C}
  {\bfseries 82} (2022) 18} [\href{https://arxiv.org/abs/2108.02383}{{\ttfamily
  2108.02383}}].

\bibitem{Cheong:2021kyc}
D.Y.~Cheong, S.M.~Lee and S.C.~Park, \emph{{Reheating in models with
  non-minimal coupling in metric and~Palatini formalisms}},
  \href{https://doi.org/10.1088/1475-7516/2022/02/029}{\emph{JCAP} {\bfseries
  02} (2022) 029} [\href{https://arxiv.org/abs/2111.00825}{{\ttfamily
  2111.00825}}].

\bibitem{Aoki:2022dzd}
S.~Aoki, H.M.~Lee, A.G.~Menkara and K.~Yamashita, \emph{{Reheating and dark
  matter freeze-in in the Higgs-R$^{2}$ inflation model}},
  \href{https://doi.org/10.1007/JHEP05(2022)121}{\emph{JHEP} {\bfseries 05}
  (2022) 121} [\href{https://arxiv.org/abs/2202.13063}{{\ttfamily
  2202.13063}}].

\bibitem{Wang:2017fuy}
Y.-C.~Wang and T.~Wang, \emph{{Primordial perturbations generated by Higgs
  field and $R^2$ operator}},
  \href{https://doi.org/10.1103/PhysRevD.96.123506}{\emph{Phys. Rev.}
  {\bfseries D96} (2017) 123506}
  [\href{https://arxiv.org/abs/1701.06636}{{\ttfamily 1701.06636}}].

\bibitem{Pi:2017gih}
S.~Pi, Y.-l.~Zhang, Q.-G.~Huang and M.~Sasaki, \emph{{Scalaron from
  $R^2$-gravity as a heavy field}},
  \href{https://doi.org/10.1088/1475-7516/2018/05/042}{\emph{JCAP} {\bfseries
  1805} (2018) 042} [\href{https://arxiv.org/abs/1712.09896}{{\ttfamily
  1712.09896}}].

\bibitem{He:2020ivk}
M.~He, R.~Jinno, K.~Kamada, A.A.~Starobinsky and J.~Yokoyama, \emph{{Occurrence
  of tachyonic preheating in the mixed Higgs-R$^2$ model}},
  \href{https://doi.org/10.1088/1475-7516/2021/01/066}{\emph{JCAP} {\bfseries
  01} (2021) 066} [\href{https://arxiv.org/abs/2007.10369}{{\ttfamily
  2007.10369}}].

\bibitem{Bezrukov:2020txg}
F.~Bezrukov and C.~Shepherd, \emph{{A heatwave affair: mixed Higgs-$R^2$
  preheating on the lattice}},
  \href{https://doi.org/10.1088/1475-7516/2020/12/028}{\emph{JCAP} {\bfseries
  12} (2020) 028} [\href{https://arxiv.org/abs/2007.10978}{{\ttfamily
  2007.10978}}].

\bibitem{He:2020qcb}
M.~He, \emph{{Perturbative Reheating in the Mixed Higgs-$R^2$ Model}},
  \href{https://doi.org/10.1088/1475-7516/2021/05/021}{\emph{JCAP} {\bfseries
  05} (2021) 021} [\href{https://arxiv.org/abs/2010.11717}{{\ttfamily
  2010.11717}}].

\bibitem{Fumagalli:2020nvq}
J.~Fumagalli, S.~Renaux-Petel and L.T.~Witkowski, \emph{{Oscillations in the
  stochastic gravitational wave background from sharp features and particle
  production during inflation}},
  \href{https://doi.org/10.1088/1475-7516/2021/08/030}{\emph{JCAP} {\bfseries
  08} (2021) 030} [\href{https://arxiv.org/abs/2012.02761}{{\ttfamily
  2012.02761}}].

\bibitem{Boutivas:2022qtl}
K.~Boutivas, I.~Dalianis, G.P.~Kodaxis and N.~Tetradis, \emph{{The effect of
  multiple features on the power spectrum in two-field inflation}},
  \href{https://arxiv.org/abs/2203.15605}{{\ttfamily 2203.15605}}.

\bibitem{Muta:1991mw}
T.~Muta and S.D.~Odintsov, \emph{{Model dependence of the nonminimal scalar
  graviton effective coupling constant in curved space-time}},
  \href{https://doi.org/10.1142/S0217732391004206}{\emph{Mod. Phys. Lett. A}
  {\bfseries 6} (1991) 3641}.

\bibitem{Elizalde:1993ee}
E.~Elizalde and S.D.~Odintsov, \emph{{Renormalization group improved effective
  potential for gauge theories in curved space-time}},
  \href{https://doi.org/10.1016/0370-2693(93)91427-O}{\emph{Phys. Lett. B}
  {\bfseries 303} (1993) 240}
  [\href{https://arxiv.org/abs/hep-th/9302074}{{\ttfamily hep-th/9302074}}].

\bibitem{Elizalde:1993ew}
E.~Elizalde and S.D.~Odintsov, \emph{{Renormalization group improved effective
  Lagrangian for interacting theories in curved space-time}},
  \href{https://doi.org/10.1016/0370-2693(94)90464-2}{\emph{Phys. Lett. B}
  {\bfseries 321} (1994) 199}
  [\href{https://arxiv.org/abs/hep-th/9311087}{{\ttfamily hep-th/9311087}}].

\bibitem{Codello:2015mba}
A.~Codello and R.K.~Jain, \emph{{On the covariant formalism of the effective
  field theory of gravity and leading order corrections}},
  \href{https://doi.org/10.1088/0264-9381/33/22/225006}{\emph{Class. Quant.
  Grav.} {\bfseries 33} (2016) 225006}
  [\href{https://arxiv.org/abs/1507.06308}{{\ttfamily 1507.06308}}].

\bibitem{Markkanen:2018bfx}
T.~Markkanen, S.~Nurmi, A.~Rajantie and S.~Stopyra, \emph{{The 1-loop effective
  potential for the Standard Model in curved spacetime}},
  \href{https://doi.org/10.1007/JHEP06(2018)040}{\emph{JHEP} {\bfseries 06}
  (2018) 040} [\href{https://arxiv.org/abs/1804.02020}{{\ttfamily
  1804.02020}}].

\bibitem{DeSimone:2008ei}
A.~De~Simone, M.P.~Hertzberg and F.~Wilczek, \emph{{Running Inflation in the
  Standard Model}},
  \href{https://doi.org/10.1016/j.physletb.2009.05.054}{\emph{Phys. Lett.}
  {\bfseries B678} (2009) 1} [\href{https://arxiv.org/abs/0812.4946}{{\ttfamily
  0812.4946}}].

\bibitem{Degrassi:2012ry}
G.~Degrassi, S.~Di~Vita, J.~Elias-Miro, J.R.~Espinosa, G.F.~Giudice, G.~Isidori
  et~al., \emph{{Higgs mass and vacuum stability in the Standard Model at
  NNLO}}, \href{https://doi.org/10.1007/JHEP08(2012)098}{\emph{JHEP} {\bfseries
  08} (2012) 098} [\href{https://arxiv.org/abs/1205.6497}{{\ttfamily
  1205.6497}}].

\bibitem{Buttazzo:2013uya}
D.~Buttazzo, G.~Degrassi, P.P.~Giardino, G.F.~Giudice, F.~Sala, A.~Salvio
  et~al., \emph{{Investigating the near-criticality of the Higgs boson}},
  \href{https://doi.org/10.1007/JHEP12(2013)089}{\emph{JHEP} {\bfseries 12}
  (2013) 089} [\href{https://arxiv.org/abs/1307.3536}{{\ttfamily 1307.3536}}].

\bibitem{Cespedes:2012hu}
S.~Cespedes, V.~Atal and G.A.~Palma, \emph{{On the importance of heavy fields
  during inflation}},
  \href{https://doi.org/10.1088/1475-7516/2012/05/008}{\emph{JCAP} {\bfseries
  05} (2012) 008} [\href{https://arxiv.org/abs/1201.4848}{{\ttfamily
  1201.4848}}].

\bibitem{Achucarro:2012yr}
A.~Achucarro, V.~Atal, S.~Cespedes, J.-O.~Gong, G.A.~Palma and S.P.~Patil,
  \emph{{Heavy fields, reduced speeds of sound and decoupling during
  inflation}}, \href{https://doi.org/10.1103/PhysRevD.86.121301}{\emph{Phys.
  Rev. D} {\bfseries 86} (2012) 121301}
  [\href{https://arxiv.org/abs/1205.0710}{{\ttfamily 1205.0710}}].

\bibitem{GrootNibbelink:2001qt}
S.~Groot~Nibbelink and B.J.W.~van Tent, \emph{{Scalar perturbations during
  multiple field slow-roll inflation}},
  \href{https://doi.org/10.1088/0264-9381/19/4/302}{\emph{Class. Quant. Grav.}
  {\bfseries 19} (2002) 613}
  [\href{https://arxiv.org/abs/hep-ph/0107272}{{\ttfamily hep-ph/0107272}}].

\bibitem{DiMarco:2002eb}
F.~Di~Marco, F.~Finelli and R.~Brandenberger, \emph{{Adiabatic and isocurvature
  perturbations for multifield generalized Einstein models}},
  \href{https://doi.org/10.1103/PhysRevD.67.063512}{\emph{Phys. Rev. D}
  {\bfseries 67} (2003) 063512}
  [\href{https://arxiv.org/abs/astro-ph/0211276}{{\ttfamily
  astro-ph/0211276}}].

\bibitem{Peterson:2011yt}
C.M.~Peterson and M.~Tegmark, \emph{{Testing multifield inflation: A geometric
  approach}}, \href{https://doi.org/10.1103/PhysRevD.87.103507}{\emph{Phys.
  Rev. D} {\bfseries 87} (2013) 103507}
  [\href{https://arxiv.org/abs/1111.0927}{{\ttfamily 1111.0927}}].

\bibitem{Greenwood:2012aj}
R.N.~Greenwood, D.I.~Kaiser and E.I.~Sfakianakis, \emph{{Multifield Dynamics of
  Higgs Inflation}},
  \href{https://doi.org/10.1103/PhysRevD.87.064021}{\emph{Phys. Rev. D}
  {\bfseries 87} (2013) 064021}
  [\href{https://arxiv.org/abs/1210.8190}{{\ttfamily 1210.8190}}].

\bibitem{Kaiser:2013sna}
D.I.~Kaiser and E.I.~Sfakianakis, \emph{{Multifield Inflation after Planck: The
  Case for Nonminimal Couplings}},
  \href{https://doi.org/10.1103/PhysRevLett.112.011302}{\emph{Phys. Rev. Lett.}
  {\bfseries 112} (2014) 011302}
  [\href{https://arxiv.org/abs/1304.0363}{{\ttfamily 1304.0363}}].

\bibitem{Mulryne:2016mzv}
D.J.~Mulryne and J.W.~Ronayne, \emph{{PyTransport: A Python package for the
  calculation of inflationary correlation functions}},
  \href{https://doi.org/10.21105/joss.00494}{\emph{J. Open Source Softw.}
  {\bfseries 3} (2018) 494} [\href{https://arxiv.org/abs/1609.00381}{{\ttfamily
  1609.00381}}].

\bibitem{Planck:2018vyg}
{\scshape Planck} collaboration, \emph{{Planck 2018 results. VI. Cosmological
  parameters}},
  \href{https://doi.org/10.1051/0004-6361/201833910}{\emph{Astron. Astrophys.}
  {\bfseries 641} (2020) A6}
  [\href{https://arxiv.org/abs/1807.06209}{{\ttfamily 1807.06209}}].

\bibitem{Planck:2018jri}
{\scshape Planck} collaboration, \emph{{Planck 2018 results. X. Constraints on
  inflation}}, \href{https://doi.org/10.1051/0004-6361/201833887}{\emph{Astron.
  Astrophys.} {\bfseries 641} (2020) A10}
  [\href{https://arxiv.org/abs/1807.06211}{{\ttfamily 1807.06211}}].

\bibitem{Bardeen:1985tr}
J.M.~Bardeen, J.R.~Bond, N.~Kaiser and A.S.~Szalay, \emph{{The Statistics of
  Peaks of Gaussian Random Fields}},
  \href{https://doi.org/10.1086/164143}{\emph{Astrophys. J.} {\bfseries 304}
  (1986) 15}.

\bibitem{Green:2004wb}
A.M.~Green, A.R.~Liddle, K.A.~Malik and M.~Sasaki, \emph{{A New calculation of
  the mass fraction of primordial black holes}},
  \href{https://doi.org/10.1103/PhysRevD.70.041502}{\emph{Phys. Rev.}
  {\bfseries D70} (2004) 041502}
  [\href{https://arxiv.org/abs/astro-ph/0403181}{{\ttfamily
  astro-ph/0403181}}].

\bibitem{Young:2014ana}
S.~Young, C.T.~Byrnes and M.~Sasaki, \emph{{Calculating the mass fraction of
  primordial black holes}},
  \href{https://doi.org/10.1088/1475-7516/2014/07/045}{\emph{JCAP} {\bfseries
  1407} (2014) 045} [\href{https://arxiv.org/abs/1405.7023}{{\ttfamily
  1405.7023}}].

\bibitem{Yoo:2018kvb}
C.-M.~Yoo, T.~Harada, J.~Garriga and K.~Kohri, \emph{{Primordial black hole
  abundance from random Gaussian curvature perturbations and a local density
  threshold}}, \href{https://doi.org/10.1093/ptep/pty120}{\emph{PTEP}
  {\bfseries 2018} (2018) 123E01}
  [\href{https://arxiv.org/abs/1805.03946}{{\ttfamily 1805.03946}}].

\bibitem{Yoo:2019pma}
C.-M.~Yoo, J.-O.~Gong and S.~Yokoyama, \emph{{Abundance of primordial black
  holes with local non-Gaussianity in peak theory}},
  \href{https://doi.org/10.1088/1475-7516/2019/09/033}{\emph{JCAP} {\bfseries
  1909} (2019) 033} [\href{https://arxiv.org/abs/1906.06790}{{\ttfamily
  1906.06790}}].

\bibitem{Young:2020xmk}
S.~Young and M.~Musso, \emph{{Application of peaks theory to the abundance of
  primordial black holes}},
  \href{https://doi.org/10.1088/1475-7516/2020/11/022}{\emph{JCAP} {\bfseries
  11} (2020) 022} [\href{https://arxiv.org/abs/2001.06469}{{\ttfamily
  2001.06469}}].

\bibitem{Yoo:2020dkz}
C.-M.~Yoo, T.~Harada, S.~Hirano and K.~Kohri, \emph{{Abundance of Primordial
  Black Holes in Peak Theory for an Arbitrary Power Spectrum}},
  \href{https://doi.org/10.1093/ptep/ptaa155}{\emph{PTEP} {\bfseries 2021}
  (2021) 013E02} [\href{https://arxiv.org/abs/2008.02425}{{\ttfamily
  2008.02425}}].

\bibitem{Wang:2021kbh}
Q.~Wang, Y.-C.~Liu, B.-Y.~Su and N.~Li, \emph{{Primordial black holes from the
  perturbations in the inflaton potential in peak theory}},
  \href{https://doi.org/10.1103/PhysRevD.104.083546}{\emph{Phys. Rev. D}
  {\bfseries 104} (2021) 083546}
  [\href{https://arxiv.org/abs/2111.10028}{{\ttfamily 2111.10028}}].

\bibitem{Harada:2013epa}
T.~Harada, C.-M.~Yoo and K.~Kohri, \emph{{Threshold of primordial black hole
  formation}}, \href{https://doi.org/10.1103/PhysRevD.88.084051,
  10.1103/PhysRevD.89.029903}{\emph{Phys. Rev.} {\bfseries D88} (2013) 084051}
  [\href{https://arxiv.org/abs/1309.4201}{{\ttfamily 1309.4201}}].

\bibitem{Harada:2015yda}
T.~Harada, C.-M.~Yoo, T.~Nakama and Y.~Koga, \emph{{Cosmological
  long-wavelength solutions and primordial black hole formation}},
  \href{https://doi.org/10.1103/PhysRevD.91.084057}{\emph{Phys. Rev. D}
  {\bfseries 91} (2015) 084057}
  [\href{https://arxiv.org/abs/1503.03934}{{\ttfamily 1503.03934}}].

\bibitem{bradley_j_kavanagh_2019_3538999}
B.J.~Kavanagh, \emph{bradkav/pbhbounds: Release version},  Nov., 2019.
\newblock 10.5281/zenodo.3538999.

\bibitem{Schmitz:2020syl}
K.~Schmitz, \emph{{New Sensitivity Curves for Gravitational-Wave Signals from
  Cosmological Phase Transitions}},
  \href{https://doi.org/10.1007/JHEP01(2021)097}{\emph{JHEP} {\bfseries 01}
  (2021) 097} [\href{https://arxiv.org/abs/2002.04615}{{\ttfamily
  2002.04615}}].

\bibitem{schmitz_kai_2020_3689582}
K.~Schmitz, \emph{{New Sensitivity Curves for Gravitational-Wave Experiments}},
   Feb., 2020.
\newblock 10.5281/zenodo.3689582.

\bibitem{Espinosa:2018eve}
J.R.~Espinosa, D.~Racco and A.~Riotto, \emph{{A Cosmological Signature of the
  SM Higgs Instability: Gravitational Waves}},
  \href{https://doi.org/10.1088/1475-7516/2018/09/012}{\emph{JCAP} {\bfseries
  09} (2018) 012} [\href{https://arxiv.org/abs/1804.07732}{{\ttfamily
  1804.07732}}].

\bibitem{Kohri:2018awv}
K.~Kohri and T.~Terada, \emph{{Semianalytic calculation of gravitational wave
  spectrum nonlinearly induced from primordial curvature perturbations}},
  \href{https://doi.org/10.1103/PhysRevD.97.123532}{\emph{Phys. Rev. D}
  {\bfseries 97} (2018) 123532}
  [\href{https://arxiv.org/abs/1804.08577}{{\ttfamily 1804.08577}}].

\bibitem{Inomata:2019yww}
K.~Inomata and T.~Terada, \emph{{Gauge Independence of Induced Gravitational
  Waves}}, \href{https://doi.org/10.1103/PhysRevD.101.023523}{\emph{Phys. Rev.
  D} {\bfseries 101} (2020) 023523}
  [\href{https://arxiv.org/abs/1912.00785}{{\ttfamily 1912.00785}}].

\bibitem{Cheong:20xxabc}
D.Y.~Cheong, M.~He, K.~Kohri and S.C.~Park, \emph{{Work in Progress}}, .

\bibitem{Renaux-Petel:2015mga}
S.~Renaux-Petel and K.~Turzy\'nski, \emph{{Geometrical Destabilization of
  Inflation}},
  \href{https://doi.org/10.1103/PhysRevLett.117.141301}{\emph{Phys. Rev. Lett.}
  {\bfseries 117} (2016) 141301}
  [\href{https://arxiv.org/abs/1510.01281}{{\ttfamily 1510.01281}}].

\bibitem{Garcia-Saenz:2018ifx}
S.~Garcia-Saenz, S.~Renaux-Petel and J.~Ronayne, \emph{{Primordial fluctuations
  and non-Gaussianities in sidetracked inflation}},
  \href{https://doi.org/10.1088/1475-7516/2018/07/057}{\emph{JCAP} {\bfseries
  07} (2018) 057} [\href{https://arxiv.org/abs/1804.11279}{{\ttfamily
  1804.11279}}].

\bibitem{Garcia-Saenz:2018vqf}
S.~Garcia-Saenz and S.~Renaux-Petel, \emph{{Flattened non-Gaussianities from
  the effective field theory of inflation with imaginary speed of sound}},
  \href{https://doi.org/10.1088/1475-7516/2018/11/005}{\emph{JCAP} {\bfseries
  11} (2018) 005} [\href{https://arxiv.org/abs/1805.12563}{{\ttfamily
  1805.12563}}].

\bibitem{Fumagalli:2019noh}
J.~Fumagalli, S.~Garcia-Saenz, L.~Pinol, S.~Renaux-Petel and J.~Ronayne,
  \emph{{Hyper-Non-Gaussianities in Inflation with Strongly Nongeodesic
  Motion}}, \href{https://doi.org/10.1103/PhysRevLett.123.201302}{\emph{Phys.
  Rev. Lett.} {\bfseries 123} (2019) 201302}
  [\href{https://arxiv.org/abs/1902.03221}{{\ttfamily 1902.03221}}].

\bibitem{Renaux-Petel:2021yxh}
S.~Renaux-Petel, \emph{{Inflation with strongly non-geodesic motion:
  theoretical motivations and observational imprints}},
  \href{https://doi.org/10.22323/1.398.0128}{\emph{PoS} {\bfseries EPS-HEP2021}
  (2022) 128} [\href{https://arxiv.org/abs/2111.00989}{{\ttfamily
  2111.00989}}].

\end{thebibliography}\endgroup

\end{document}